\documentclass[12pt,reqno]{amsart}
\usepackage{amsmath}
\usepackage{amsxtra}
\usepackage{amscd}
\usepackage{amsthm}
\usepackage{amsfonts}
\usepackage{amssymb}
\usepackage{eucal}
\usepackage{cases}
\usepackage{paralist}
\usepackage{array}
\usepackage{etex}
\usepackage{color}
\usepackage{float}
\usepackage{cite}

\usepackage{graphicx,color}

\allowdisplaybreaks

\def\smallskip{\vskip\smallskipamount}
\def\medskip{\vskip\medskipamount}
\def\bigskip{\vskip\bigskipamount}

\def\({\left(}
\def\){\right)}
\def\[{\left[}
\def\]{\right]}
\newcommand{\ds}[1]{\displaystyle #1}

\newcommand{\nn}{\nonumber}
\newcommand{\bea}{\begin{eqnarray}}
	\newcommand{\ena}{\end{eqnarray}}
\def\bel{\begin{eqnarray}}
	\def\enl{\end{eqnarray}}

\def\ba{\begin{eqnarray}}
\def\ea{\end{eqnarray}}

\def\lb{\label}
\def\be{\begin{equation}}
\def\ee{\end{equation}}

\newcommand{\sltr}{\mathfrak{sl}_3}
\newcommand{\slth}{\widehat{\mathfrak{sl}}_2}

\newcommand{\slnhat}{\widehat{\mathfrak{sl}}_n}
\newcommand{\res}{{\rm res}}

\newenvironment{tenumerate}{
	\begin{enumerate}
		
	}{\end{enumerate}}
\newcommand{\bi}{\begin{tenumerate}}
	\newcommand{\ei}{\end{tenumerate}}
\newcommand{\isoto}[1][]%
{{\mathop{\buildrel{\sim}\over\longrightarrow}\limits_{#1}}}
\newcommand{\To}[1][\phantom{aaaa}]{\xrightarrow{\,#1\,}}

  \let\la=\lambda

\def\2{\frac{1}{2}} \def\4{\frac{1}{4}}

\def\6{\partial}

\def\+{\dagger}

\def\<{\langle} \def\>{\rangle}

\def\ctg{\, {\rm ctg}\,}

\renewcommand{\cot}{\ctg}

\numberwithin{equation}{section}

\begin{document}
	
\begin{title}[]
  {Flat connection on four-dimensional lattice,
  related matrix difference equations and their solutions}
\end{title}
\date{\today}
\author{H.~Boos, A.P.~Isaev}
\address{Physics Department, University of Wuppertal, D-42097,
	Wuppertal, Germany}\email{hboos@uni-wuppertal.de}
\address{Bogoliubov Laboratory of Theoretical Physics, JINR, Dubna,
	141 980 Moscow Region, Russia}\email{isaevap@thsun1.jinr.ru}

\begin{abstract}
In the paper \cite{BHN} the reduced density matrix of the $\sltr$-invariant fundamental exchange model was calculated for the operator length up to
three by means of the reduced quantum Knizhnik-Zamolodchikov equation.	
In this paper we present the solution of some special difference problem originated from the study of the reduced density matrix
for the operator length 4. This difference problem is related to a four-dimensional zero-curvature condition and has a clear geometrical meaning were we have a trivial
fiber bundle $\mathbb{CP}^{3}\times \mathbb{C}^4$ with a vector
function which takes value in $\mathbb{C}^4$ and the base being
the projective space $\mathbb{CP}^{3}$. The local connection
coefficients satisfy the above mentioned zero-curvature or flatness
condition. The solution we discuss here is given in terms
of the $\Gamma$-function, its logarithmic derivative,
hypergeometric function	and some other related functions
defined via the functional relations of difference type.
\end{abstract}

\maketitle

\section{\bf Introduction}

The difference equations appear in many
 areas of mathematics
 and mathematical physics.
 We can mention difference equations that appear in the study of
some nonlinear integrable classical differential equations like KdV,
KP, Toda lattice equations, 
Painlev\'{e} equations etc. (see \cite{Hiet}, \cite{Nijh2}, \cite{BobSur}
 and references therein).

 The most relevant area in our case is the problem of the
exact computation of correlation functions for integrable models
both of statistical mechanics and quantum field theory.
 In the case of quantum statistical models it is, for example,
the Heisenberg XXZ spin chain based on the quantum group $U_q(\slth)$,
and its generalizations related to the quantum groups of the higher
rank, for instance, $U_q(\slnhat)$.
One of the most powerful methods here is related to the study of solutions quantum Knizhnik-Zamolodchikov (qKZ) equation or its variant --
the so-called reduced quantum
Knizhnik-Zamolodchikov equation (rqKZ) which was introduced
in \cite{BKS}, \cite{BJMST} (see also \cite{BHN}). We stress
 that both  qKZ and rqKZ equations are examples of
difference equations.


Among the difference equations the worth highlighting problem is
 the matrix linear difference equations and
  their analytic theory \cite{Birk}, \cite{Bor},
 \cite{Krich}. The matrix linear difference problem
 can be written as
 \be
 \lb{psiA}
 \Psi(\lambda + 1) = A(\lambda) \; \Psi(\lambda)  \; ,
 \ee
 where $\lambda \in \mathbb{C}$ is a parameter,
 $\Psi(\lambda)$ is unknown $M \times K$ matrix and $M \times M$
 matrix $A(\lambda)$ is called connection and
 has poles of the first order
 at the finite set of points $\lambda_j$, $j=1,2,...,N$:
 \be
 \lb{psiA1}
 A(\lambda) = \sum_{j=1}^{N}
\frac{A_j}{\lambda - \lambda_j}  \; .
 \ee
 It is obvious that
equation (\ref{psiA}) is a discrete analog of the linear
system of differential equations
 \be
 \lb{psiA2}
 \frac{d}{d \lambda} \Psi(\lambda) =  A(\lambda) \Psi(\lambda) \; ,
 \ee
which arises in the context of the
 Riemann-Hilbert (inverse monodromy) problem.
 One can formulate this problem as following. \\
{\it For an arbitrary $N$  points $\lambda_j \in \mathbb{C}$,
construct $M \times M$ matrix function $\Psi(\lambda)$
 such that: \\
1) 	$\Psi(\infty) = I_M$, \\
2) elements of $\Psi(\lambda)$ are holomorphic
 for all $\lambda \neq \lambda_j$, $\forall j$, \\
3) elements of $\Psi(\lambda)$ has regular singular points at
 $\lambda \neq \lambda_j$,
$j = 1, . . . ,N$, with given monodromy
matrices $M_j$.} \\
 Then in the case when the monodromy matrices $M_j$ are
independent of the parameters $\lambda_i$, one can search
the function $\Psi(\lambda) \in {\rm Mat}_M(\mathbb{C})$ as
a solution of the matrix differential equation (\ref{psiA2})
 with matrix $A$ given in (\ref{psiA1}), where
$SL(M,\mathbb{C})$-valued coefficients $A_j$ obey the system of
Schlesinger equations \cite{Schl}
 \be
 \lb{psiA3}
 \frac{\partial}{\partial \lambda_i} A_j =
 \frac{[A_i, \; A_j ]}{\lambda_i - \lambda_j} \; , \;\;\;
\lambda_i \neq \lambda_j \; , \;\;\;
\frac{\partial}{\partial \lambda_i} A_i =
 - \sum_{j\neq i} \frac{[A_i, \; A_j ]}{\lambda_i - \lambda_j}
 \; .
 \ee
  One can consider
  (see \cite{JiMiw02}
  )
  this system as an integrable hierarchy
  with $\lambda_i$ interpreted as a set of times.
  The integrals of motion of the system (\ref{psiA3})
 are obviously given by the eigenvalues of matrices $A_j$
 and $\tau$-function was introduced in \cite{JiMiw02}.
 Remarkably,
  in special case of $M=2$ and special choice
  of poles $\lambda_i$, the system (\ref{psiA3})
 reduces to Painlev\'{e} equations (it was shown
 in \cite{JiMiw01}, \cite{JiMiw02}, 
 ;
see also \cite{ItsNov}, \cite{KitKor} and references therein).
 Then the isomonodromy problem for difference analog (\ref{psiA})
of the linear matrix equation (\ref{psiA2}) yields the
 lattice versions of Painlev\'{e}
equations (see e.g. \cite{Kit}).
 All these studies  began with the paper
 \cite{JiMiw01} where it was shown that
 correlation functions and n-particle reduced density matrices
  of certain quantum Bose system (described by
  non-linear Schr\"{e}dinger equation) satisfies a
 system of non-linear partial differential equations
 of Painlev\'{e} type. Derivation of
these equations was based on the monodromy preserving deformation
theory related to the problem (\ref{psiA2}).

\vspace{0.2cm}


In our paper we are going to discuss solution to one concrete matrix
equation of difference type which stems from the problem
of calculation of the correlation functions of the rational $\sltr$-invariant
model in case of the operator length four. The work on this problem
is still in progress.
It contains several parts of different degree of difficulty. We decided to
describe in this paper one of this parts which seems to be the most
nontrivial and essential
for the solution of the whole problem: namely to find density matrix
in case of the operators of length 4 because we believe that it is of
a separate interest.

The paper is organized as follows. In Sect. 2 we formulate the matrix
equation (\ref{G}) of difference type. One can consider
this matrix difference equation as definition of flat connection
on a lattice, or as discrete analog of auxiliary
 linear problem for special integrable system.
 Then, in Sect. 2 we reduce the matrix equation
  to the scalar difference equation (\ref{g}), which is solved by means of
 simpler methods. In Sect. 3 we
 find four independent nontrivial solutions of the
 scalar equation (\ref{g}) and in Sect. 4 we reconstruct the
 matrix solution $G$ of equation (\ref{G}). Finally in Sect. 5
 we investigate the features of the matrix $G$. In particular we
  calculate inverse matrix $G^{-1}$, which is much simpler then matrix $G$
  and is needed for evaluation of correlation functions
  for operators of length 4 in rational $s\ell(3)$-model
  (see \cite{BHN}).

\section{Formulation of the problem}

More precisely we consider the following matrix equation of difference type:
\bea
G(\la_1|\la_2,\la_3,\la_4)\;A(\la_1|\la_2,\la_3,\la_4) = G(\la_1-3|\la_2,\la_3,\la_4)
\label{G}
\ena
with $4\times 4$ matrices $G$ and $A$ which depend on four spectral parameters $\la_1,\cdots,\la_4$
with entries that are complitely symmetric wrt the parameters $\la_2,\la_3,\la_4$. We can also call (\ref{G}) {\it a difference pure gauge
condition} where $A$ is a kind of connection and $G$ is a gauge transform.
The matrix $A$ is a known rational matrix
\bea
A(\la_1|\la_2,\la_3,\la_4)=
\frac{M(\la_1|\la_2,\la_3,\la_4)}{\prod\limits_{j=2}^4
(\la_{1j}-3)(\la_{1j}-1)(\la_{1j}+1)},\quad \la_{1j}:=\la_1-\la_j
\;\;\;\;\; (j=2,3,4) \; ,
\label{A}
\ena
where the matrix $M$ has polynomial entries. Unfortunately, some of these entries are given by rather bulky polynomials. So, we believe that
it would be not so instructive just to show an explicit form of them.
We will rather try to define the matrix $A$ in a less direct way.
We start to discuss some properties of the matrix
 $A$ in the next subsection.

 \vspace{0.1cm}

 \noindent
{\bf Remark 1.} Matrix difference equation (\ref{G}) possesses
the gauge symmetry
$$
\begin{array}{c}
G(\la_1|\la_2,\la_3,\la_4)\; \to \;
G(\la_1|\la_2,\la_3,\la_4) \; R(\la_1|\la_2,\la_3,\la_4)
\, , \\ [0.2cm]
A(\la_1|\la_2,\la_3,\la_4)  \; \to \;
R(\la_1|\la_2,\la_3,\la_4)^{-1} \;
A(\la_1|\la_2,\la_3,\la_4) \; R(\la_1-3|\la_2,\la_3,\la_4) \; ,
\end{array}
$$
where $R(\la_1|\la_2,\la_3,\la_4)$ is a gauge matrix.

\vspace{0.1cm}

\noindent
{\bf Remark 2.}
	We consider the case when solution $G(\la_1|\la_2,\la_3,\la_4)$ and matrix $M(\la_1|\la_2,\la_3,\la_4)$
in difference equation (\ref{G}) with connection
(\ref{A}), depend only on the differences $\la_{1j}:=\la_1-\la_j$.
If we relax in (\ref{G}) the symmetry under all permutations of
three parameters $\lambda_2,\lambda_3,\lambda_4$ and take
 $x:=\la_{12}$, $t:=\la_{2} -\la_3$, $z := \la_{3} -\la_4$
 as independent variables,
 then matrix difference equation (\ref{G}) can be interpreted
  as auxiliary spectral problem for difference integrable model,
  where $x$, $t$ and $z$ are respectively coordinate,
  time and spectral parameter.

\subsection{Zero curvature condition}

	The remarkable fact is that the matrix (\ref{A})
satisfies two equations. The first equation is
 \bea
&&	A(\la_1|\la_2,\la_3,\la_4)\; A(\la_2|\la_1-3,\la_3,\la_4)\;
=\;A(\la_2|\la_1,\la_3,\la_4)\; A(\la_1|\la_2-3,\la_3,\la_4)\label{zerocurv}
\ena
and we call this equation as `zero curvature' condition
(see explanations below). The second equation is
\bea
 A(\la_1|\la_2,\la_3,\la_4)\, A(\la_2|\la_1-3,\la_3,\la_4)\,
 A(\la_3|\la_1-3,\la_2-3,\la_4)\, A(\la_4|\la_1-3,\la_2-3,\la_3-3)=\mathbb{I}	\nn\\
\label{AAAA}
\ena
	where $\mathbb{I}$ is unit $4 \times 4$ matrix.

	To clarify the geometric
origin of equations (\ref{zerocurv}) and (\ref{AAAA})
we consider a four component complex valued vector, say, $\Psi$ which
depends on the above four parameters $\la_1,\la_2,\la_3,\la_4$:
\bea
\Psi = \Psi(\la_1,\la_2,\la_3,\la_4)
\label{Psi}
\ena
and is invariant under translation
\bea
\Psi(\la_1,\la_2,\la_3,\la_4) =
 \Psi(\la_1+3,\la_2+3,\la_3+3,\la_4+3)
\label{translPsi}
\ena
where we consider $\la_i$ as coordinates of a point
$\mathbf{x}=(\la_1,\la_2,\la_3,\la_4)$ of a four-dimensional
 space. In fact, because of the above translation invariance  we are dealing
with vectors $\Psi$ defined on the four-dimensional torus.
 The matrix $A$ acting on the vector $\Psi$ (for some reasons
to the 
	right
)
 makes a discrete transfer with step 3 of this vector from the original point
$\mathbf{x}=(\la_1,\la_2,\la_3,\la_4)$ along the first axis to the point $\mathbf{x'}=(\la_1-3,\la_2,\la_3,\la_4)$:
\bea
\Psi(\la_1,\la_2,\la_3,\la_4)A(\la_1|\la_2,\la_3,\la_4) =
\Psi(\la_1-3,\la_2,\la_3,\la_4)
\label{transfer1}
\ena
Similarly, we have
\bea
\Psi(\la_1,\la_2,\la_3,\la_4)A(\la_2|\la_1,\la_3,\la_4) =
\Psi(\la_1,\la_2-3,\la_3,\la_4)
\label{transfer2}
\ena
for the transfer along the second axis and so on. We can go alone
the first axis and then along the second one or vice versa coming to the
same result
\bea
&&\Psi(\la_1,\la_2,\la_3,\la_4)\;A(\la_1|\la_2,\la_3,\la_4)\;
A(\la_2|\la_1-3,\la_3,\la_4) \nn\\
=&&
\Psi(\la_1,\la_2,\la_3,\la_4)\;A(\la_2|\la_1,\la_3,\la_4)\;
A(\la_1|\la_2-3,\la_3,\la_4)\; =\;
\Psi(\la_1-3,\la_2-3,\la_3,\la_4)
\label{transfer12}
\ena
This happens to be true for any vector $\Psi$
 due to the `zero curvature' condition (\ref{zerocurv})
and as pointed out above, the matrix $A$ can be thought as a
	4-dimensional connection with components
 $$
 (A_1,A_2,A_3,A_4) = \bigl(A(\la_1|\la_2,\la_3,\la_4), \,
 A(\la_2|\la_1,\la_3,\la_4), \,
  A(\la_3|\la_1,\la_2,\la_4), \, A(\la_4|\la_1,\la_2,\la_3) \bigr) \; .
 $$
 Here we take into account the symmetry of the function
 $A(\la_1|\la_2,\la_3,\la_4)$ with respect to all permutations of
 three variables $\lambda_2,\lambda_3,\lambda_4$. Each component $A_j$
is responsible for transfer of
the vector $\Psi$ arguments with step 3 along a related
 direction $j$. If we go alone
the axis $j$ and then along the axis $k$ or vice versa
 we obtain `zero curvature' condition (\ref{zerocurv})
 with substitution $1 \leftrightarrow j$ and $2 \leftrightarrow k$.

Also we can go along the first axis then along the second, third axis and finally along the fourth axis coming to the vector
$\Psi(\la_1-3,\la_2-3,\la_3-3,\la_4-3)$ which is equal to the original
vector $\Psi(\la_1,\la_2,\la_3,\la_4)$ due to the translational invariance
(\ref{translPsi}). So, in this way we return to the starting point.
The corresponding relation looks as (\ref{AAAA}).

\noindent
{\bf Remark 3.} 
	Comparing equations
 (\ref{G}) and (\ref{transfer1}), one can assume
 that as a vector $\Psi$ it is possible to take a row of matrix $G$.
 But this is not the case, since in general
 the rows of matrix $G$ does not satisfy
 the second equation (\ref{transfer2}) etc.

\subsection{Degeneration poins}

We have a simple formula for the determinant of the matrix $M$:
\bea
\det{\Bigl(M(\la_1|\la_2,\la_3,\la_4)\Bigr)}=
\prod\limits_{j=2}^4
(\la_{1,j}-4)^2(\la_{1,j}-3)(\la_{1,j}-2)^3
(\la_{1,j}-1)\la_{1,j}^3(\la_{1,j}+1)^2
\label{detM}
\ena
The degeneration points, where the above determinant becomes zero,
correspond to the following situation with the rank of the matrix $M$:
\bea
&&
	\text{rank}=1: \quad \la_1=\la_j,\;\la_1=\la_j+2,\quad j=2,3,4\nn\\
&&	\text{rank}=2: \quad \la_1=\la_j-1,\;\la_1=\la_j+4,\nn\\
&&
	\text{rank}=3: \quad \la_1=\la_j+1, \quad \la_1=\la_j+3
\label{degen}
\ena
Using the zero curvature condition (\ref{zerocurv}), we come
to the relations:
\bea
&&
M(\la_2+2|\la_2,\la_3,\la_4)\; M(\la_2|\la_2-1,\la_3,\la_4)=0,\quad
M(\la_2+1|\la_2,\la_3,\la_4)\; M(\la_2|\la_2-2,\la_3,\la_4)=0\nn\\
&&
M(\la_2+3|\la_2,\la_3,\la_4)\; M(\la_2|\la_2,\la_3,\la_4)=0,\quad\quad
M(\la_2|\la_2,\la_3,\la_4)\; M(\la_2|\la_2-3,\la_3,\la_4)=0\nn\\
&&
M(\la_2+4|\la_2,\la_3,\la_4)\; M(\la_2|\la_2+1,\la_3,\la_4)=0,\quad
M(\la_2-1|\la_2,\la_3,\la_4)\; M(\la_2|\la_2-4,\la_3,\la_4)=0\nn\\
\label{ker}
\ena
	for degeneration points on the hyper-plane $\la_1=\la_2$. Note
 that there are similar relations for
degeneration points on the hyper-plane $\la_1=\la_3$ (or $\la_1=\la_4$).


 Now, for the matrix $A$ we can write down the following formula:
\bea
A(\la_1|\la_2,\la_3,\la_4)=\sum\limits_{i=2}^4
\Bigl(\frac{M^{(3)}_i}{\la_{1,i}-3}
+\frac{M^{(1)}_i}{\la_{1,i}-1}+
\frac{M^{(-1)}_i}{\la_{1,i}+1}\Bigr)+
\mathbb{I}+M_-(\la_1|\la_2,\la_3,\la_4)  \; ,
\label{formA}
\ena
where the residues $M_i^{(a)}, a=3, 1, -1$ are matrix functions of
$\la_2,\la_3,\la_4$. They are proportional to the matrices
$M(\la_i+3|\la_i,\la_j,\la_k), M(\la_i+1|\la_i,\la_j,\la_k),
M(\la_i-1|\la_i,\la_j,\la_k)$ respectively where  $(i,j,k)$ stands for
any permutation of $(2,3,4)$. Hence, for example, we get from
(\ref{ker}):
\bea
&&
M(\la_2+3|\la_2+3,\la_3,\la_4)\; M^{(3)}_2=0,
\quad\quad \text{rank}(M^{(3)}_i)=3, \quad i=2,3,4\quad\quad
\nn\\
&&
M(\la_2+3|\la_2+1,\la_3,\la_4)\; M^{(1)}_2=0,
\quad\quad \text{rank}(M^{(1)}_i)=3,
\nn\\
&&
M(\la_2+3|\la_2-1,\la_3,\la_4)\; M^{(-1)}_2=0,
\quad\quad \text{rank}(M^{(-1)}_i)=2 \; .
\nn
\ena
The matrix $M_-$ in (\ref{formA}) is lower-triangular with
polynomial entries
\bea
M_-(\la_1|\la_2,\la_3,\la_4)=
\begin{pmatrix}
		0 & 0 & 0 & 0 \\
		0 & 0 & 0 & 0 \\
		\star  & 0 & 0 & 0\\
		\star  & \star & \star & 0
\end{pmatrix} \; .
\label{M-}
\ena
We will give explicit expressions for the elements of the matrix residues
$M_i^{(3)}, M_i^{(\pm 1)}$ and the matrix $M_-$ in Appendix B.

Below we shall use the notation
$$
\la:=\la_{12},\quad \mu:=\la_{13},\quad \nu:=\la_{14} \; .
$$
In terms of these parameters the formulae look a bit simpler.
	 We take into account that the matrices $A$ and $G$ depend only on
parameters $\lambda_{1j}$.
Thus, with slight abuse of notation
we will write
$$
A\equiv A(\la,\mu,\nu),\quad G\equiv G(\la,\mu,\nu) \; ,
$$
and the equation (\ref{G}) becomes
\bea
G(\la,\mu,\nu)\;A(\la,\mu,\nu) = G(\la-3,\mu-3,\nu-3) \; ,
\label{eqG}
\ena
	where entries of matrices $A(\la,\mu,\nu)$ and $G(\la,\mu,\nu)$
are symmetric functions of $\la,\mu,\nu$.

\subsection{From matrix difference relation to a scalar difference equation}

Our main goal is to explicitly find a {\it non-singular} $4\times4$ matrix $G$
which satisfies the above matrix difference equation (\ref{eqG}).
Let $g_{i,j},\; i,j=1,\cdots 4$ be the entries of the matrix $G$.
Then we can proceed as follows: we choose one row of the matrix $G$, for example, the first one and take the same row in the equation (\ref{eqG}).
In this way we get four closed difference equations for the elements
of this row $g_{1,1},g_{1,2},g_{1,3},g_{1,4}$. Then it turns out to be
more convenient to solve the third and the fourth equations first and express the functions $g_{1,1}$ and $g_{1,2}$ through $g_{1,3}$ and $g_{1,4}$.
The residual two equations will contain linear combinations
of $g_{1,i}(\la,\mu,\nu), g_{1,i}(\la-3,\mu-3,\nu-3), g_{1,i}(\la-6,\mu-6,\nu-6), \quad i=3,4$. By taking resultants we get
a single linear difference equation for one function $g_{1,4}$ which contains
$g_{1,4}(\la+3j,\mu+3j,\nu+3j)$ where $j$ is running from $(-3)$ to $2$
with extremely huge polynomial coefficients.
Let us for simplicity call the unknown function just $g$
$$
g_{1,4}(\la,\mu,\nu)\equiv g(\la,\mu,\nu)
$$
As appeared one can find two rather bulky polynomials
$B_1(\la,\mu,\nu)$ and $B_2(\la,\mu,\nu)$ such that the above linear
equation has a form
$$
B_1(\la,\mu,\nu)\cdot {\text{Eq}}(\la,\mu,\nu)+
B_2(\la,\mu,\nu)\cdot {\text{Eq}}(\la-3,\mu-3,\nu-3)=0
$$
where the equation Eq looks:
\bea
{\text{Eq}}(\la,\mu,\nu):=\sum_{j=-2}^2 C_{3j}(\la,\mu,\nu)\;g(\la+3j,\mu+3j,\nu+3j)=0
\label{g}
\ena
with some known fully-symmetric polynomial coefficients $C_{3j}(\la,\mu,\nu),\; j=-2,\cdots, 2$. Remarkable fact is that
these coefficients are relatively simple in comparison with
the above mentioned huge coefficients and polynomails $B_1,B_2$.
We describe the coefficients $C_{3j}$ 
	(more precisely their
linear combinations  (\ref{SC}))
in Subsection {\bf \ref{coeS}}
and in Appendix.

For further consideration we need the following Definition. \\

{\bf Definition.} {\it By a quasi-constant we call a function
$c(\la,\mu,\nu)$ which is invariant under transformation of coordinates:
\bea
(\la,\mu,\nu) \;\; \To \;\; (\la+3,\mu+3,\nu+3) \; ,
\label{shift3}
\ena
i.e., a quasi-constant is a function  that satisfies:}
\bea
c(\la,\mu,\nu)=c(\la+3,\mu+3,\nu+3).
\label{quasiconst}
\ena
We expect that the equation (\ref{g}) has four solutions $g_1,\cdots, g_4$ which are linear independent modulo quasi-constants.
So, by linear-independence modulo quasi-constants
we mean that there are no quasi-constants
$c_i, \; i=1,\cdots, 4$ such that
$$
\sum_{i=1}^4 c_i(\la,\mu,\nu)\;g_i(\la,\mu,\nu) = 0
$$
We associate these four functions $g_i$ with four entries of the matrix $G$ corresponding to the
last column: $g_{i,4}\equiv g_i$.

As appeared\footnote{
	It follows from asymptotic
consideration of solutions of eqs. (\ref{g})
and (\ref{f}).}, all solutions $g_i$ of equation (\ref{g})
have the form:
\bea
g_i(\la,\mu,\nu)=u(\la,\mu,\nu)\cdot \Delta(f_i)(\la,\mu,\nu)
\label{gi}
\ena
where the factor $u$ is given by:
\bea
u(\la,\mu,\nu):=(\la-\la^{-1})(\mu-\mu^{-1})(\nu-\nu^{-1})
\label{u}
\ena
and $\Delta$ is a second order difference operator:
\bea
\Delta(f) (\la,\mu,\nu):= f(\la+3,\mu+3,\nu+3)\;-\;2f(\la,\mu,\nu)\; +\;f(\la-3,\mu-3,\nu-3)
\label{Delta}
\ena

Below we will need the first order operators as well:
\bea
{\ds\Delta_{\pm}(f) (\la,\mu,\nu):= f(\la+\frac32,\mu+\frac32,\nu+\frac32)\;\pm\; f(\la-\frac32,\mu-\frac32,\nu-\frac32)}
\label{Deltapm}
\ena
The operator $\Delta$ is then:
\bea
\Delta=\Delta_-^2
\label{Delta1}
\ena

The above four functions $f_i$ are non-trivial solutions of the equation
\bea
\sum_{j=-3}^3 S_{3j}(\la,\mu,\nu)\; f(\la+3j,\mu+3j,\nu+3j)=0
\label{f}
\ena
with some fully-symmetric polynomials $S_{3j}(\la,\mu,\nu),\; j=-3,\cdots,3$
of degree of homogeneity 19 that are connected with the above coefficients $C_{3j}(\la,\mu,\nu)$ and satisfy two simple relations
\bea
\sum_{j=-3}^3 S_{3j}(\la,\mu,\nu)=\sum_{j=-3}^3 S_{3j}(\la,\mu,\nu)\;j=0
\label{S}
\ena
which correspond to the two trivial solutions of the equation (\ref{f}):
\bea
f^{(a)}(\la,\mu,\nu)=1, \quad f^{(b)}(\la,\mu,\nu)=\la+\mu+\nu
\label{triv}
\ena
Of course, both these solutions belong to the kernel of the above
operator $\Delta$
$$
\Delta(f^{(a)})=\Delta(f^{(b)})=0
$$
Explicitly, the coefficients $S_{3j}$ are related to the coefficients
$C_{3j}$ as follows:
\bea
&&
S_9 = \bar{C}_6,\quad
S_6=-2\bar{C}_6+\bar{C}_3,\nn\\
&&
S_3=\bar{C}_6-2\bar{C}_3+\bar{C}_0,\quad
S_0=\bar{C}_3-2\bar{C}_0+\bar{C}_{-3},\quad
S_{-3}=\bar{C}_{-6}-2\bar{C}_{-3}+\bar{C}_0\label{SC}\\
&&
S_{-9} = \bar{C}_{-6},\quad
S_{-6}=-2\bar{C}_{-6}+\bar{C}_{-3}
\nn
\ena
where
\bea
\bar{C}_{3j} =
n_{3j}\cdot C_{3j}
\label{barC}
\ena
with the multiplier
$$
n_{3j}=\frac{u(\la+3j,\;\mu+3j,\;\nu+3j)}{(\la^2-1)(\la^2-4)\;
	(\mu^2-1)(\mu^2-4)\;(\nu^2-1)(\nu^2-4)}
$$
where the factor $u$ is given by (\ref{u}).

Our goal is to find the non-trivial solutions of the equation (\ref{f}).

\subsection{\bf Coefficients $\bf S_{3j}$\label{coeS}}

Here we discuss some additional to (\ref{S})
properties of coefficients $S_{3j}$.
First of all, the coefficients $S_{3j}(\lambda,\mu,\nu)$
are completely symmetric functions of $\lambda,\mu,\nu$
 that satisfy
\bea
S_{3j}(\lambda,\mu,\nu) = - S_{-3j}(-\lambda,-\mu,-\nu)  \; .
\label{asymS}
\ena
It means that only $S_{9},S_{6},S_{3}$ and $S_{0}$
are independent functions.

It is known that any symmetric polynomial of variables
$\lambda,\mu,\nu$  can
be represented as a polynomial of basic symmetric functions
\bea
s_1 = \la + \mu + \nu,\quad
s_2 = \la \mu  + \lambda \nu + \mu \nu,\quad
s_3 = \la\, \mu \, \nu  \; .
\label{sypol}
\ena
Below we give explicit formulas for coefficients
$S_{9},S_{6}$ and $S_{3}$ in terms of
basic symmetric polynomials $s_1$, $s_2$ and $s_3$
(coefficient $S_0$ is expressed via $S_{9},S_{6}$ and $S_{3}$
by means of (\ref{S})).

\subsubsection{\bf \em Coefficient $\bf S_9$}

For $S_9$ we have representation
\bea
&&
S_9 = (\lambda_{(4)} \mu_{(4)} \nu_{(4)}) (\lambda_{(5)}
\mu_{(5)} \nu_{(5)}) (\lambda_{(7)} \mu_{(7)}  \nu_{(7)}) \,
\bar{S}_9,\label{s9}
\\
&&
\bar{S}_9=
\bar{S}_9^{(0)} + \bar{S}_9^{(1)} + \dots + \bar{S}_9^{(10)}
\nn
\ena
where $\bar{S}_9^{(k)}$ -- symmetric $k$-th order polynomial
of $\mu,\nu,\lambda$. Here and below we use concise notation
$\la_{(a)}:=\la+a$, $\mu_{(a)}:=\mu+a$,
$\nu_{(a)}:=\nu+a$.

In addition to (\ref{asymS}),
 the coefficient $\bar{S}_9$ in (\ref{s9})
possesses reflection symmetry
\begin{equation}
\label{S9m}
\bar{S}_9(\la,\mu,\nu)=\bar{S}_9(3-\la,3-\mu,3-\nu).
\end{equation}

In terms of basis of symmetric polynomials
(\ref{sypol}) the homogeneous coefficients $\bar{S}_9^{(k)}$
in (\ref{s9}) are
$$
\bar{S}_9^{(0)}= -18504 \; , \;\;\;
\bar{S}_9^{(1)}  = -720 \, s_1  \; , \;\;\;
\bar{S}_9^{(2)}  = -6390 s_2 +2240 \, s_1^2  \; ,
$$
$$
\bar{S}_9^{(3)}  = 4599 \, s_3-2862 s_1 s_2+384 \, s_1^3 \; , \;\;\;                                     \bar{S}_9^{(4)}  = 2158 \, s_1^2 s_2 -2543 \, s_1 \, s_3
-2392 \, s_2^2 -264 \, s_1^4 \; ,
$$
$$
\bar{S}_9^{(5)}  = -3132 s_2 \, s_3 + 60 s_1 s_2^2+849 \, s_1^2 s_3
+6 s_1^3 s_2 \; ,
$$
$$
\bar{S}_9^{(6)}  = -2352 s_3^2+2066 s_1 s_2 s_3 -110 s_2^3
- 76 s_1^2 \, s_2^2-321 s_1^3 \, s_3 \; ,
$$
$$
\bar{S}_9^{(7)}  = 1035 s_3^2 s_1
-549 s_2^2 s_3 -42 s_1^2 s_2 s_3+42 s_1 s_2^3
\; ,
$$
$$
\bar{S}_9^{(8)}  = -4 s_2^4-120 s_3^2 s_2 -183 s_3^2 s_1^2
+ 85s_1s_2^2 s_3 \; ,
$$
\begin{equation}
\label{coef1}
\bar{S}_9^{(9)}  = 9 s_3 (9 s_3^2-2 s_2^3  + 5 s_1 s_2 s_3)
\; , \;\;\;
\bar{S}_9^{(10)}  = 6 s_3^2 ( s_2^2 -3 s_3 s_1)  \; .
\end{equation}
Thus, taking into account prefactor (see (\ref{s9}))
$$
\begin{array}{c}
(\lambda_4 \mu_4 \nu_4) (\lambda_5
\mu_5 \nu_5) (\lambda_7 \mu_7  \nu_7) = s_3^3 + 16 s_3^2 s_2 +83 s_2^2 s_3 +90 s_3^2 s_1
+532 s_3^2+140 s_2 ^3+ \\ [0.2cm]
908 s_1  s_2  s_3 +5230 s_2  s_3
+2240 s_2^2 s_1 +11620 s_1 ^2 s_2 +12600 s_2^2+2409 s_1^2 s_3
+26924 s_1  s_3 + \\ [0.2cm]
+127120 s_1  s_2  +72827  s_3
+19600 s_1 ^3+337260 s_2 +313600 s_1^2+1626800 s_1 +2744000 \; ,
\end{array}
$$
in (\ref{s9})
we see that polynomial $S_9$ has maximal order equal to $19$.
 To find all possible homogeneous terms
$s_1^{n_1} s_2^{n_2} s_3^{n_3}$ in coefficients (\ref{coef1}),
it is useful to
associate all these terms with the Young diagram

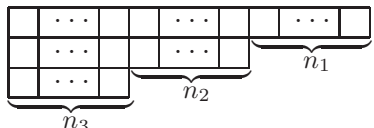
\unitlength=4mm
\begin{picture}(25,4)(-3,0.8)

\put(11,4){\line(1,0){12}}
\put(11,1){\line(1,0){4}}


\put(12.5,3.5){$\dots$}
\put(12.5,1.5){$\dots$}
\put(12.5,2.5){$\dots$}

\put(15,1){\line(0,1){3}}
\put(11,1){\line(0,1){3}}

\put(14,1){\line(0,1){3}}
\put(12,1){\line(0,1){3}}

\put(11,2){\line(1,0){4}}
\put(11,3){\line(1,0){8}}

\put(11,1){\tiny $\underbrace{\;\;\;\;\;\;\;\;\;\;\;\;
		\;\;\;\;\;\;\;\;}$}
\put(12.8,0.0){\footnotesize $n_3$}


\put(15,2){\line(1,0){4}}
\put(15,3){\line(1,0){8}}

\put(15,2){\line(0,1){2}}
\put(16,2){\line(0,1){2}}
\put(18,2){\line(0,1){2}}
\put(19,2){\line(0,1){2}}
\put(20,3){\line(0,1){1}}
\put(22,3){\line(0,1){1}}
\put(23,3){\line(0,1){1}}

\put(16.5,2.5){$\dots$}
\put(16.5,3.5){$\dots$}
\put(20.5,3.5){$\dots$}

\put(15.1,2){\tiny $\underbrace{\;\;\;\;\;\;\;\;\;\;\;\;
		\;\;\;\;\;\;\;}$}
\put(16.8,1.0){\footnotesize $n_2$}

\put(19.1,3){\tiny $\underbrace{\;\;\;\;\;\;\;\;\;\;\;\;
		\;\;\;\;\;\;\;}$}
\put(20.8,2.0){\footnotesize $n_1$}

\end{picture}

\vspace{0.3cm}

\noindent
{\bf Remark 4.} Define symmetric functions
\bea
p_2=\frac{1}{2} \Bigl( (\lambda-\mu)^2 + (\lambda-\nu)^2 +
(\mu-\nu)^2 \Bigr)=s_1^2-3s_2  \; ,
\label{p2}
\ena
\begin{equation}
\label{tozd4}
\begin{array}{c}
p_3' = (\lambda+\nu-2\mu)
(\lambda+\mu- 2\nu)(\nu+\mu-2 \lambda)=
9 s_2 s_1 -2 s_1^3 -27 s_3 \; ,
\end{array}
\end{equation}
that are invariant under simultaneous shift
 \begin{equation}
\label{geninv}
\mu \to \mu +a \; , \;\;\; \nu \to \nu +a \; ,
\;\;\; \lambda \to \lambda +a \;\;\;\;\;\;\;\; (\forall \; a) \; .
\end{equation}
 	 According to the definition
 of quasi-constants, as functions which are invariant
 under transformation  (\ref{shift3}),
the functions (\ref{tozd4})
 are polynomial quasi-constants.
  We call symmetric polynomials invariant under general shift (\ref{geninv})
  as symmetric polynomial quasi-constants.
Any symmetric polynomial quasi-constant is a function of
$p_2$ and  $p_3'$ only. In other words, the algebra of
symmetric polynomial quasi-constants has only two
generators $p_2$ and  $p_3'$.
As an example consider
polynomial quasi-constants
\begin{equation}
\label{tozd5}
\begin{array}{c}
p_4 = (\lambda-\mu)^4 + (\lambda-\nu)^4 + (\mu-\nu)^4
\; ,  \\ [0.2cm]
p_4' = (\lambda-\mu)^2 (\lambda-\nu)^2 + (\mu-\nu)^2 (\lambda-\nu)^2 + (\lambda-\mu)^2  (\mu-\nu)^2 \; , \\ [0.2cm]
p_6 = (\lambda-\mu)^6 + (\lambda-\nu)^6 + (\mu-\nu)^6 \; , \\ [0.2cm]
p_6' = (\lambda-\mu)^2(\lambda-\nu)^2(\mu-\nu)^2  \; .
\end{array}
\end{equation}
For these polynomials we have
\begin{equation}
\label{tozd7}
p_4 = 2 \, p_2^2  \; , \;\;\;  p_4' =p_2^2 \; , \;\;\;
p_6' = \frac{1}{27} \bigl( 4  \, p_2^3 - (p_3')^2 \bigr)  \; , \;\;\;
p_6 = 2 \, p_2^3 + 3 p_6' \; .
\end{equation}

	Introduce new basic symmetric polynomials
 (cf. (\ref{sypol}), (\ref{tozd4}))
\begin{equation}
\label{sypol2}
\sigma_1 = \frac{1}{6} s_1 - \frac{3}{4}  \; , \;\;\;
\bar{p}_2 = \frac{1}{3} p_2 - \frac{1}{4} \; ,  \;\;\;
p_3 =  \frac{1}{27} p_3 ' \; .
\end{equation}
Under the change of variables $\mu \to 3 - \mu$,
$\nu \to 3 - \nu$, $\lambda \to 3 - \lambda$, these
 basic symmetric polynomials are transformed as following
\begin{equation}
\label{trans1}
\sigma_1 \to - \sigma_1 \; , \;\;\;
\bar{p}_2 \to \bar{p}_2 \; , \;\;\; p_3 \to - p_3   \; .
\end{equation}
In terms of variables (\ref{sypol2}) the coefficient
$\bar{S}_9$ (according to formulas (\ref{s9}), (\ref{coef1}))
is written as
\begin{equation}
\label{sypol4}
\begin{array}{c}
\bar{S}_9 = (4608 \bar{p}_2-8064)\sigma_1^8+6912 p_3\sigma_1^7+
(39480-26112 \bar{p}_2-1920 \bar{p}_2^2)\sigma_1^6+
(-4608 \bar{p}_2-27792) p_3\sigma_1^5+ \\ [0.2cm]
(-1728 p_3^2-48972+96 \bar{p}_2^3+5424 \bar{p}_2^2+43302 \bar{p}_2)\sigma_1^4
+(624 \bar{p}_2^2+32592+4440 \bar{p}_2) p_3\sigma_1^3 \\ [0.2cm]
+(24 \bar{p}_2^4-\frac{61323}{2} \bar{p}_2+(-2070+504 \bar{p}_2) p_3^2
-648 \bar{p}_2^3+\frac{50295}{2}-\frac{1515}{2} \bar{p}_2^2)\sigma_1^2
\\ [0.2cm]
+(2775 \bar{p}_2+24 \bar{p}_2^3-1065 \bar{p}_2^2-3639
+108 p_3^2)p_3 \sigma_1 \\ [0.2cm]
- \frac{35}{2} \bar{p}_2^4+(-\frac{1221}{2}- \frac{321}{2}
\bar{p}_2+6 \bar{p}_2^2) p_3^2
+\frac{3521}{8} \bar{p}_2^3-21924-\frac{26075}{8} \bar{p}_2^2+
\frac{25893}{2} \bar{p}_2 \; .
\end{array}
\end{equation}
Taking into account this representation,
the invariance (\ref{S9m}) of $\bar{S}_9$
becomes manifest, since the function  (\ref{sypol4})  is a
polynomial of variables $\sigma_1^{2k}$, $(p_3 \cdot \sigma_1)$,
$\bar{p}_2$ and $p_3^{2k}$ which are obviously invariant under the
transformations (\ref{trans1}).


\subsubsection{\bf Coefficient $\bf S_6$.}

For the coefficient $S_6$ we also have factorized form
\bea
&&
S_6 = -2 \, (\lambda_{(4)} \mu_{(4)} \nu_{(4)})  \,\bar{S}_6,
\label{S6}\\
&&
\bar{S}_6=\bar{S}_6^{(0)} + \bar{S}_6^{(1)} + \dots + \bar{S}_6^{(16)}  \nn
\ena
and for homogeneous
symmetric polynomials $\bar{S}_6^{(k)}$ we have explicit expressions
 in terms of basic symmetric polynomials (\ref{sypol})
 which we give in the Appendix A.

\subsubsection{\bf Coefficient $\bf S_3$.}
For coefficient $S_3$ we have expansion
\begin{equation}
\label{s3}
S_3 = \bar{S}_3^{(0)} + \bar{S}_3^{(1)} + \dots + \bar{S}_3^{(19)}  \; ,
\end{equation}
where homogeneous symmetric polynomials $\bar{S}_3^{(k)}$
with degree of homogeneity $k$ are explicitly given in
terms of polynomials $s_1,s_2,s_3$ in Appendix A.

\subsubsection{\bf Coefficient $\bf S_0$.}
Coefficient $S_0$ can be expressed with the help of
first relation in (\ref{S})
via coefficients $S_3$, $S_6$ and $S_9$
and we will not present the explicit form of $S_0$ here.

\section{\bf Non-trivial solutions}

\subsection{\bf First solution}

	As was mentioned above, the equation (\ref{f}) has two
trivial solutions (\ref{triv}).
The first non-trivial solution of (\ref{f})
(and the most simple one) which we call $f^{(0)}$ appeared to be
\bea
f^{(0)}(\la,\mu,\nu)= \frac{\Gamma(\frac{2+\lambda}{3})}{\Gamma(\frac{1+\lambda}{3})} \cdot
\frac{\Gamma(\frac{2+\mu}{3})}{\Gamma(\frac{1+\mu}{3})} \cdot
\frac{\Gamma(\frac{2+\nu}{3})}{\Gamma(\frac{1+\nu}{3})} \; .
\label{f0}
\ena
It has rather simple transformation property
\bea
f^{(0)}(\la+3,\mu+3,\nu+3)=\gamma_3(\la,\mu,\nu)\;f^{(0)}(\la,\mu,\nu)
\label{transformf0}
\ena
where
\bea
\gamma_3(\la,\mu,\nu)=
\frac{\la_{(2)}\mu_{(2)}\nu_{(2)}}{\la_{(1)}\mu_{(1)}\nu_{(1)}}
\label{gamma3}
\ena
	We also introduce rational functions
 $\gamma_{3j}(\la,\mu,\nu)$ such that
\bea
f^{(0)}(\la+3j,\mu+3j,\nu+3j)=
\gamma_{3j}(\la,\mu,\nu)\;f^{(0)}(\la,\mu,\nu) \; ,
\label{trans3j}
\ena
and for functions $\gamma_{3j}$
we have the recurrent relations and identities
\bea
 \gamma_{3j+3}(\la,\mu,\nu) & = & \gamma_3(\la+3j,\mu+3j,\nu+3j)
 \; \gamma_{3j}(\la,\mu,\nu) \nn\\
& = & \gamma_{3j}(\la+3,\mu+3,\nu+3) \; \gamma_3(\la,\mu,\nu) \; ,
\nn\\
\gamma_{-3j}(\la,\mu,\nu) & = & \gamma_{3j}(-\la,-\mu,-\nu) =
 [ \gamma_{3j}(\la-3j,\mu-3j,\nu-3j)]^{-1} \; , \nn\\
\gamma_0(\la,\mu,\nu) & = & 1  \; .
\label{relationgamma}
\ena
We can easily check that $f^{(0)}$ given by (\ref{f0}) is indeed a solution
of (\ref{f}) 
	by using the relation (\ref{trans3j})
 and 
the following property of the coefficients $S_{3j}$:
\bea
\sum_{j=-3}^3 S_{3j}\;\gamma_{3j}(\la,\mu,\nu)=0
\label{Sgamma}
\ena

\subsection{Second solution}

The second solution of eq. (\ref{f}) which we call $f^{(1)}$ transforms
inhomogeneously:
\bea
f^{(1)}(\la+3,\mu+3,\nu+3)=\gamma_3(\la,\mu,\nu)\;f^{(1)}(\la,\mu,\nu)-
\beta_3(\la,\mu,\nu)  \; ,
\label{transformf1}
\ena
where
\bea
\beta_3(\la,\mu,\nu):=\frac{1}{\la_{(1)}\mu_{(1)}\nu_{(1)}}  \; .
\label{bet3}
\ena
 	From eq. (\ref{transformf1}) one can deduce a general formula
 \bea
f^{(1)}(\la+3j,\mu+3j,\nu+3j)=\gamma_{3j}(\la,\mu,\nu)\;f^{(1)}(\la,\mu,\nu)-
\beta_{3j}(\la,\mu,\nu) \; ,
\label{tran3j1}
\ena
where the functions $\beta_{3j}$ are again
defined
from the recurrent relations
\bea
&&\beta_{3j+3}(\la,\mu,\nu)=\gamma_3(\la+3j,\mu+3j,\nu+3j)
\beta_{3j}(\la,\mu,\nu)+\beta_3(\la+3j,\mu+3j,\nu+3j)\nn\\
&&\beta_{-3j}(\la,\mu,\nu)=\beta_{3j}(-\la,-\mu,-\nu)\nn\\
&&\beta_0(\la,\mu,\nu)=0
\label{relationbeta}
\ena
The validity of the solution $f^{(1)}$ follows from
	the formula (\ref{tran3j1}) and
further property of
the coefficients $S_{3j}$
\bea
\sum_{j=-3}^3 S_{3j}\;\beta_{3j}=0  \; ,
\label{Sbeta}
\ena
which can 
	be verified directly. By using (\ref{gamma3}),
 (\ref{trans3j}), (\ref{bet3}) and (\ref{relationbeta})
we write (\ref{tran3j1}) in the form
 \bea
\frac{f^{(1)}(\la+3j,\mu+3j,\nu+3j)}{f^{(0)}(\la+3j,\mu+3j,\nu+3j)}=
\frac{f^{(1)}(\la,\mu,\nu)}{f^{(0)}(\la,\mu,\nu)} -
\sum_{k=0}^{j-1}\Gamma_k(\la,\mu,\nu),
\label{tran3j2}
\ena
 where
 \bea
\Gamma_k(\la,\mu,\nu)=\frac{1}{27}\cdot \frac{\Gamma(\frac{1+\la}{3}+k)\Gamma(\frac{1+\mu}{3}+k)\Gamma(\frac{1+\nu}{3}+k)}
{\Gamma(\frac{5+\la}{3}+k)\Gamma(\frac{5+\mu}{3}+k)
\Gamma(\frac{5+\nu}{3}+k)}  \; .
\label{Gammak}
\ena

Of course, the equation (\ref{tran3j2}) does not define the
solution $f^{(1)}/f^{(0)}$ uniquely since one can always add
 to $f^{(1)}/f^{(0)}$ arbitrary quasi-constant $c$.
 So, modulo the addition of
$c \cdot f^{(0)}$ we find the following new solution:
\bea
&&
f^{(1)}(\la,\mu,\nu)= f^{(0)}(\la,\mu,\nu)\cdot\sum_{k=0}^{\infty}\Gamma_k(\la,\mu,\nu),
\label{f1}
\ena
 with $\Gamma_k$ defined in (\ref{Gammak}).
We can write this in terms of the hypergeometric function
\bea
&&
f^{(1)}(\la,\mu,\nu)=\frac{1}{(2+\la)(2+\mu)(2+\nu)}\;
{}_4F_3\Biggl(
\begin{array}{ccccc}
	{1}	& \frac{1+\la}3 & \frac{1+\mu}3 &\frac{1+\nu}3|\\
	 & \quad & \quad & \quad \;\;|&1\\
	&{\;}\frac{5+\la}3 & \frac{5+\mu}3 & \frac{5+\nu}3 |
\end{array}
\Biggr)
\label{f1a}
\ena

\noindent
{\bf Remark 5.}
Due to the symmetry property  (\ref{asymS}) 
we realize that $f^{(0)}(-\la,-\mu,-\nu)$ and $f^{(1)}(-\la,-\mu,-\nu)$
are the solutions of the equation (\ref{f}) also but they are not
linear-independent in the above sense because they satisfy the same
transformation properties (\ref{transformf0},\ref{transformf1}) as for $f^{(0)}(\la,\mu,\nu)$ and $f^{(1)}(\la,\mu,\nu)$. In particular, we
observe that
\bea
f^{(0)}(-\la,-\mu,-\nu)=C^{(0)}(\la,\mu,\nu)\cdot f^{(0)}(\la,\mu,\nu)
\label{f0minus}
\ena
where
$$
C^{(0)}(\la,\mu,\nu)=\frac{\sin{\pi(\frac{1-\la}3)}}{\sin{\pi(\frac{1+\la}3)}}
\cdot \frac{\sin{\pi(\frac{1-\mu}3)}}{\sin{\pi(\frac{1+\mu}3)}}\cdot
\frac{\sin{\pi(\frac{1-\nu}3)}}{\sin{\pi(\frac{1+\nu}3)}}
$$
is a quasi-constant.

\subsection{Search for two further solutions}

In order to find further solutions to the equation (\ref{f}) let us
slightly generalize the transformation law (\ref{transformf1})
\bea
f^{(2)}(\la+3,\mu+3,\nu+3)=\gamma_3(\la,\mu,\nu)\;f^{(2)}(\la,\mu,\nu)-
\beta_3(\la,\mu,\nu)h^{(1)}(\la,\mu,\nu)
\label{transformf2}
\ena
where we should find such a function $h^{(1)}(\la,\mu,\nu)$ which
transforms:
\bea
h^{(1)}(\la+3,\mu+3,\nu+3) = h^{(1)}(\la,\mu,\nu) + h^{(2)}(\la+3,\mu+3,\nu+3)
\label{transformh1}
\ena
If we set $h^{(1)}=1, h^{(2)}=0$ we come to the solution $f^{(1)}$.

Let us try the function $f^{(2)}$ as a solution of the equation (\ref{f}).
If we use the transformation properties (\ref{transformf2},\ref{transformh1})
we observe that the contribution of the functions $f^{(2)}$ and $h^{(1)}$
drops out.
So, we come to the equation for the function $h^{(2)}$:
\bea
\sum_{j=-2}^2 \tilde{S}_{3j}(\la,\mu,\nu)\; h^{(2)}(\la+3j,\mu+3j,\nu+3j)=0
\label{h2}
\ena
where the coefficients $\tilde{S}_{3j}$ are related to the
above coefficients $S_{3j}$ as follows:
\bea
&&\tilde{S}_{6}(\la,\mu,\nu)=
-\beta_3(\la+6,\mu+6,\nu+6)\cdot S_9(\la,\mu,\nu)
\label{tildeS6}\\
&&\tilde{S}_{3}(\la,\mu,\nu)=
-\beta_6(\la+3,\mu+3,\nu+3)\cdot S_9(\la,\mu,\nu)
-\beta_3(\la+3,\mu+3,\nu+3)\cdot S_6(\la,\mu,\nu)
\nn
\\
&&\tilde{S}_{0}(\la,\mu,\nu)=
-\beta_9(\la,\mu,\nu)\cdot S_9(\la,\mu,\nu)
-\beta_6(\la,\mu,\nu)\cdot S_6(\la,\mu,\nu)
-\beta_3(\la,\mu,\nu)\cdot S_3(\la,\mu,\nu)
\nn
\\
&&
\tilde{S}_{-3j}(\la,\mu,\nu)=\tilde{S}_{3j}(-\la,-\mu,-\nu)
\label{symtildeS}
\ena

An important observation is that the coefficients $\tilde{S}_{3j}$ satisfy
the property similar to (\ref{S}):
\bea
\sum_{j=-2}^2 \tilde{S}_{3j}(\la,\mu,\nu)=\sum_{j=-2}^2 \tilde{S}_{3j}(\la,\mu,\nu)\;j=0
\label{tildeS}
\ena
It means that there are again two trivial solutions like (\ref{triv})
\bea
h^{(2,a)}(\la,\mu,\nu)=1, \quad h^{(2,b)}(\la,\mu,\nu)=\la+\mu+\nu
\label{trivh2}
\ena
which lie in the kernel of the operator $\Delta$.
In order to find two non-trivial solutions let us introduce the function
\bea
\sigma(\la,\mu,\nu):=\Delta(h^{(2)})(\la,\mu,\nu)
\label{sigma}
\ena
which satisfies the equation
\bea
\sum_{j=-1}^1 S^{(\sigma)}_{3j}(\la,\mu,\nu)\; \sigma(\la+3j,\mu+3j,\nu+3j)=0
\label{eqsigma}
\ena
where
\bea
&&
S^{(\sigma)}_{3}(\la,\mu,\nu)=\tilde{S}_6(\la,\mu,\nu),\quad
S^{(\sigma)}_{0}(\la,\mu,\nu)=2\tilde{S}_6(\la,\mu,\nu)+\tilde{S}_3(\la,\mu,\nu),\quad
S^{(\sigma)}_{-3}(\la,\mu,\nu)=\tilde{S}_{-6}(\la,\mu,\nu)\nn\\
\label{Ssigma}
\ena
It turns out that the equation (\ref{eqsigma}) has two {\it rational}
solutions
\bea
\sigma(\la,\mu,\nu)=\sigma_{\pm}(\la,\mu,\nu),\quad
\sigma_{\pm}(-\la,-\mu,-\nu)=\pm \sigma_{\pm}(\la,\mu,\nu)
\label{sigmapm}
\ena
Explicitly we have:
\bea
&&
\sigma_{+}=(3-p_2)\cdot \biggl(\frac{1}{\la_{(1)}\la_{(2)}\mu_{(1)}\mu_{(2)}\nu_{(1)}\nu_{(2)}}+
\frac{1}{\la_{(-1)}\la_{(-2)}\mu_{(-1)}\mu_{(-2)}\nu_{(-1)}\nu_{(-2)}}\biggr)\nn\\
&&
\ds{-\,\frac{2\, s_1+15}{3\,\la_{(1)}\mu_{(1)}\nu_{(1)}}\;
-\,\frac{2\, s_1-15}{3\,\la_{(-1)}\mu_{(-1)}\nu_{(-1)}}\;
+\,\frac{2\, s_1+21}{3\,\la_{(2)}\mu_{(2)}\nu_{(2)}}\;
+\,\frac{2\, s_1-21}{3\,\la_{(-2)}\mu_{(-2)}\nu_{(-2)}}
}
\label{sigma+}\\
&&\nn\\
&&\nn\\
&&
\sigma_{-}=
-\frac{(2s_2+9s_1+32)(2p_2+3s_2+6s_1+6)}
{3\,\la_{(1)}\la_{(2)}\mu_{(1)}\mu_{(2)}\nu_{(1)}\nu_{(2)}}
+\frac{(2s_2-9s_1+32)(2p_2+3s_2-6s_1+6)}
{3\,\la_{(-1)}\la_{(-2)}\mu_{(-1)}\mu_{(-2)}\nu_{(-1)}\nu_{(-2)}}\nn\\
&&
+\,\frac{4 s_1+27}{\la_{(2)}\mu_{(2)}\nu_{(2)}}\;
-\,\frac{4 s_1-27}{\la_{(-2)}\mu_{(-2)}\nu_{(-2)}}
\label{sigma-}
\ena
where the polynomials $s_1,s_2,p_2$ are defined in (\ref{sypol},\ref{p2}).

Now we have to ``integrate'' twice these expressions in order to get
an explicit form of the function $h^{(2)}$ using the formula
(\ref{sigma}).
So, for the symmetric and anti-symmetric cases $h^{(2)}\equiv h_{\pm}^{(2)}$ we should solve the equation
\bea
\Delta\bigl(h_{\pm}^{(2)}\bigr)=\sigma_{\pm}
\label{eqh2p}
\ena
The result looks as follows
\bea
h_{\pm}^{(2)}(\la,\mu,\nu) =&
c_{\pm}(\la|\mu,\nu)\cdot\psi(\frac{1+\la}3)\;\pm\; c_{\pm}(-\la|-\mu,-\nu)\cdot\psi(\frac{2+\la}3)\nn\\
+&c_{\pm}(\mu|\la,\nu)\cdot\psi(\frac{1+\mu}3)\;\pm\; c_{\pm}(-\mu|-\la,-\nu)\cdot\psi(\frac{2+\mu}3)\nn\\
+&c_{\pm}(\nu|\la,\mu)\cdot\psi(\frac{1+\nu}3)\;\pm\; c_{\pm}(-\nu|-\la,-\mu)\cdot\psi(\frac{2+\nu}3)
\label{h2pm}
\ena
where $\psi(x)=\partial_x\log{\Gamma(x)}$ is a standard notation for the logarithmic derivative of $\Gamma$-function and the coefficient $c_+$ is
\bea
&&
c_+(\la|\mu,\nu)=\frac{(\la-2)}3\cdot c_1(\la|\mu,\nu)-c_2(\la|\mu,\nu)
\;=\;
\frac19\cdot \frac{(2\la-\mu-\nu)(2\la^2-4\la+1-\mu-\nu-2\mu\nu)}
{(\la-\mu)(\la-\mu-1)(\la-\nu)(\la-\nu-1)}\nn\\
&&
\label{c+}\\
&&
c_-(\la|\mu,\nu)=
r(\la|\mu,\nu)\cdot c_+(\la|\mu,\nu)
\label{c-}\\
\nn\\
&&
r(\la|\mu,\nu):=\frac{5(\la-\mu)(\la-\nu)+2(\mu-\nu)^2-3}{2\la-\mu-\nu}
\label{r}
\ena
with two quasi-constants $c_1,c_2$
\bea
&&
c_1(\la|\mu,\nu)=\frac23\cdot
\frac{(2\la-\mu-\nu)(2\la-\mu-\nu-3)}{(\la-\mu)(\la-\mu-1)(\la-\nu)(\la-\nu-1)}
\nn\\
&&
c_2(\la|\mu,\nu)=\frac19\cdot
\frac{(2\la-\mu-\nu)\bigl(2(\la-\mu-5/2)(\la-\nu-5/2)-3/2\bigr)}
{(\la-\mu)(\la-\mu-1)(\la-\nu)(\la-\nu-1)}
\label{c12}
\ena


Below we will also need the quasi-constants $c'_1,c'_2$ that give
\bea
&&
c_+(-\la|-\mu,-\nu)=\frac{(\la-1)}3\cdot c'_1(\la|\mu,\nu)-c'_2(\la|\mu,\nu)
\nn\\
&&
c'_1(\la|\mu,\nu)=c_1(-\la|-\mu,-\nu),\nn\\
&&
c'_2(\la|\mu,\nu)=-\frac19\cdot
\frac{(2\la-\mu-\nu)\bigl(2(\la-\mu-1/2)(\la-\nu-1/2)-15/2\bigr)}
{(\la-\mu)(\la-\mu+1)(\la-\nu)(\la-\nu+1)}
\label{c'12}
\ena

\noindent
{\bf Remark 6.}
Of course, the expressions (\ref{h2pm}) for the functions $h_{\pm}^{(2)}$ given by
(\ref{h2pm}) can be generalized.
Due to the relation (\ref{tildeS}) one can always add the element
of the kernel of operator $D$ like $c(\la,\mu,\nu)+(\la+\mu+\nu)d(\la,\mu,\nu)$
with quasi-constants $c,d$. Using this property, we could write down
the result (\ref{h2pm}) in a more symmetric form if we take into account
the identity:
$$
\psi\Bigl(\frac{2+\la}3\Bigr)= \psi\Bigl(\frac{1-\la}3\Bigr)+\pi \cot{\Bigl(\frac{\pi}3(1-\la)\Bigr)}
$$
Since the second term in the r.h.s. is
a quasi-constant, we could substitute $\psi(\frac{2+\la}3)\rightarrow
\psi(\frac{1-\la}3)$ in (\ref{h2pm}).
The equation (\ref{h2}) would be still satisfied due to the evident relation
$$
\sum_{j=-2}^2 \tilde{S}_{3j} \cdot c_+(-\la-3j|-\mu-3j,-\nu-3j)=0.
$$
The result would be explicitly symmetric under the transform
$\la,\mu,\nu\rightarrow -\la,-\mu,-\nu$  for
$h_{+}^{(2)}$ and anti-symmetric for $h_{-}^{(2)}$. Unfortunately, in this case we would get quadratic poles in the final result for $f^{(2)}$. Hence, we will not undertake this substitution and stay with the formula (\ref{h2pm}).

The next step is to get the function $h^{(1)}$ from the equation
(\ref{transformh1}) by one more ``integration''.
The result for the symmetric  and anti-symmetric case $h^{(1)}\rightarrow h_{\pm}^{(1)}$ looks:
\bea
h_{\pm}^{(1)}(\la,\mu,\nu)=
&
b_{1,\pm}(\la|\mu,\nu)\cdot\psi(\frac{1+\la}3)\;+
\;b_{2,\pm}(\la|\mu,\nu)\cdot\psi(\frac{2+\la}3)\nn\\
+&
b_{1,\pm}(\mu|\la,\nu)\cdot\psi(\frac{1+\mu}3)\;+
\;b_{2,\pm}(\mu|\la,\nu)\cdot\psi(\frac{2+\mu}3)\nn\\
+&
b_{1,\pm}(\nu|\la,\mu)\cdot \psi(\frac{1+\nu}3)\;+\;b_{2,\pm}(\nu|\la,\mu)\cdot\psi(\frac{2+\nu}3)
\label{h1+}
\ena
where the coefficients
\bea
&&
b_{1,+}(\la|\mu,\nu)=\frac{\la+1}3\cdot \biggl(\frac{\la-2}6\cdot c_1(\la|\mu,\nu)-
c_2(\la|\mu,\nu)\biggr)\nn\\
&&=\frac{(\la+1)(2\la-\mu-\nu)\bigl(\la(\mu+\nu)-2\mu\nu+3(\la-\mu-\nu)-5\bigr)}{27(\la-\mu)(\la-\mu-1)(\la-\nu)(\la-\nu-1)}
\nn\\
&&\nn\\
&&
b_{2,+}(\la|\mu,\nu)=\frac{\la+2}3\cdot \biggl(\frac{\la-1}6\cdot c'_1(\la|\mu,\nu)-
c'_2(\la|\mu,\nu)\biggr)\nn\\
&&=
-\frac{(\la+2)(2\la-\mu-\nu)\bigl(\la(\mu+\nu)-2\mu\nu+3 \la+4\bigr)}{27(\la-\mu)(\la-\mu+1)(\la-\nu)(\la-\nu+1)}\nn\\
\nn\\
&&
b_{i,-}(\la|\mu,\nu)=r(\la|\mu,\nu)
\cdot b_{i,+}(\la|\mu,\nu),\quad i=1,2
\label{b1b2+-}
\ena
with the function $r$ defined in (\ref{r}).

Interesting properties of the above coefficients are that
\bea
&&
b_{1,+}(\la|\mu,\nu)+b_{2,+}(-\la|-\mu,-\nu)=c_+(\la|\mu,\nu)
\label{prop1}\\
&&
b_{1,+}(\la|\mu,\nu)+b_{2,+}(-\la-3|-\mu-3,-\nu-3)=0
\label{prop2}
\ena

Now let us proceed to the last step, namely, to the solving
the equation (\ref{transformf2}). To this end let us rewrite it
by multiplying both sides by $1/f^{(0)}(\la,\mu,\nu)$. After a simple
algebra we arrive at 
:
\bea
\frac{f^{(2)}(\la+3,\mu+3,\nu+3)}{f^{(0)}(\la+3,\mu+3,\nu+3)}=
\frac{f^{(2)}(\la,\mu,\nu)}{f^{(0)}(\la,\mu,\nu)}\;-\;
\frac1{27}\;\frac{\Gamma(\frac{1+\la}3)
	\Gamma(\frac{1+\mu}3)\Gamma(\frac{1+\nu}3)}{\Gamma(\frac{5+\la}3)
	\Gamma(\frac{5+\mu}3)\Gamma(\frac{5+\nu}3)}\cdot h^{(1)}(\la,\mu,\nu)
\label{eqf2}
\ena
Its solution looks
\bea
&&
f^{(2)}(\la,\mu,\nu)=
f^{(0)}(\la,\mu,\nu)\cdot\sum_{k=0}^{\infty}\Gamma_k(\la,\mu,\nu)
\cdot
h^{(1)}(\la+3k,\mu+3k,\nu+3k)
\label{f2}
\ena
where the function $\Gamma_k(\la,\mu,\nu)$ was defined in (\ref{Gammak}).

For the above two solutions $h^{(1)}_{\pm}$ it can be explicitly written
down as follows:
\bea
f^{(2)}_{\pm}(\la,\mu,\nu)=
f^{(0)}(\la,\mu,\nu)\cdot\Bigl(F_{\pm}(\la|\mu,\nu)+F_{\pm}(\mu|\la,\nu)+
F_{\pm}(\nu|\la,\mu)\Bigr)
\label{f2pm}
\ena
where
\bea
&&
F_+(\la|\mu,\nu)=c_1(\la|\mu,\nu)\cdot F_{1,1}(\la|\mu,\nu)-
c_2(\la|\mu,\nu)\cdot F_{1,2}(\la|\mu,\nu)\nn\\
&&
\qquad\qquad\quad
+
c'_1(\la|\mu,\nu)\cdot F_{2,1}(\la|\mu,\nu)-
c'_2(\la|\mu,\nu)\cdot F_{2,2}(\la|\mu,\nu)\nn\\
\nn\\
&&
F_-(\la|\mu,\nu)=r(\la|\mu,\nu)\cdot F_+(\la|\mu,\nu)
\label{Fpm}
\ena
\bea
&&
F_{1,1}(\la|\mu,\nu):=\sum_{k=0}^{\infty}\;\Gamma_k(\la,\mu,\nu)\cdot
\frac12\;\Bigl(\frac{1+\la}3+k\Bigr)\Bigl(\frac{-2+\la}3+k\Bigr)\cdot \psi\Bigl(\frac{1+\la}3+k\Bigr)\nn\\
&&
F_{1,2}(\la|\mu,\nu):=\sum_{k=0}^{\infty}\;\Gamma_k(\la,\mu,\nu)\cdot
\Bigl(\frac{1+\la}3+k\Bigr)\cdot \psi\Bigl(\frac{1+\la}3+k\Bigr)\nn\\
\label{F}\\
&&
F_{2,1}(\la|\mu,\nu):=\sum_{k=0}^{\infty}\;\Gamma_k(\la,\mu,\nu)\cdot
\frac12\;\Bigl(\frac{2+\la}3+k\Bigr)\Bigl(\frac{-1+\la}3+k\Bigr)\cdot \psi\Bigl(\frac{2+\la}3+k\Bigr)\nn\\
&&
F_{2,2}(\la|\mu,\nu):=\sum_{k=0}^{\infty}\;\Gamma_k(\la,\mu,\nu)\cdot
\Bigl(\frac{2+\la}3+k\Bigr)\cdot \psi\Bigl(\frac{2+\la}3+k\Bigr)\nn\\
&&
\nn
\ena
and the quasi-constants $c_1,c_2,c'_1,c'_2$ are defined in (\ref{c12},\ref{c'12}).

\noindent
{\bf Remark 7.}
Since the behaviour at $k\to\infty$
$$
\Gamma_k(\la,\mu,\nu)\simeq k^{-4},\quad
k^2\psi(a+k)\simeq k^2\log{k}
$$
the above four sums in (\ref{F}) are perfectly convergent.\\

\noindent
{\bf Remark 8.}
One of our previous attempts to solve the equation (\ref{f}) was
to consider the pole structure of a solution. We looked for the pole
part w.r.t. the variable $\la$ in the following form:
\bea
f_{pole}(\la|\mu,\nu)=\sum_{j=-\infty}^{\infty} \frac{d_j(\mu,\nu)}{\la-1+3j}
\label{pole}
\ena
Another series of poles at $\la=1-3j$ could be analysed in the same way.
We conjectured that the residue at $j=0$ is zero: $d_0(\mu,\nu)=0$.
Then we realized that the residues must satisfy the following
functional relation:
\bea
&&
d_{j+1}(\mu,\nu)-\frac{(3j+1)}{3j} \frac{(\mu-1)(\nu-1)}{(\mu-2)(\nu-2)}\cdot
d_j(\mu-3,\nu-3)\label{relres}\\
&&
=j\Bigl(1-\frac{(3j-2)}{3j} \frac{(\mu-1)(\nu-1)}{(\mu-2)(\nu-2)}\Bigr)\cdot Y(\mu-3(j-1),\nu-3(j-1)),
\quad
j=\pm1,\pm2,\pm3\cdots
\nn
\ena
where $Y$ is an arbitrary function. If we consider analytic continuation
\bea
d_j(\mu,\nu):=\tilde{g}(3j-1|\mu,\nu)
\label{dj}
\ena
with the function $\tilde{g}(\la|\mu,\nu)$ which
fulfills a generalized equation (\ref{relres}) where we substitute
the integer $j$ by $\frac{\la-2}3$. This equation can be explicitly
solved:
\bea
&&\tilde{g}(\la|\mu,\nu)=\Bigl(\la-2+ f^{(0)}(\la,\mu,\nu)\cdot
\sum_{k=0}^{\infty}\tilde{\Gamma}_k(\la|\mu,\nu)\Bigr)\cdot \tilde{Y}(\mu-\la,\nu-\la)
\label{tildeg}\\
&&
\tilde{\Gamma}_k(\la|\mu,\nu):=\frac{\Gamma(\frac{1+\la}3+k)
	\Gamma(\frac{1+\mu}3+k)\Gamma(\frac{1+\nu}3+k)}
{\Gamma(\frac{5+\la}3+k)
	\Gamma(\frac{2+\mu}3+k)\Gamma(\frac{2+\nu}3+k)}
\ena
with some quasi-constant $\tilde{Y}(\mu-\la,\nu-\la)
$ related to the function $Y$ in (\ref{relres}).
This solution is not unique since we could add here the function
$f^{(0)}$ multiplied
by a quasi-constant because it doesn't change the r.h.s. of (\ref{relres}).
Then we realized that the function $\tilde{g}$ itself turned out to be solution
of the original equation (\ref{f}). But in fact, it is linear-dependent on the
above two solutions $f^{(0)}$ and $f^{(1)}$ given by (\ref{f0}) and (\ref{f1})
respectively.
An interesting question is how the residues of the solution (\ref{f2pm})
for $f^{(2)}_{\pm}$ are related to the above residues $d_j(\mu,\nu)$
given by (\ref{dj}). It appeared to be simpler to answer this question
for the case of negative $j=-l<0$ since we have an identity:
\bea
&&{\ds\res_{\la=-1-3l} \Bigl( f^{(2)}_{\pm}(\la,\mu,\nu)\Bigr)=
\res_{\la=-1-3l} \Bigl(f^{(0)}(\la,\mu,\nu)F_{\pm}(\la|\mu,\nu)\Bigr)
=\tilde{g}(-1-3l|\mu,\nu)\cdot Y_{\pm}(-1-3l|\mu,\nu)}\nn\\
&&
\label{resFpm}
\ena
with quasi-constants:
\bea
&&
Y_+(\la|\mu,\nu):=-\frac{2\la-\mu-\nu}{9(\la-\mu)(\la-\mu-1)
	(\la-\nu)(\la-\nu-1)},\nn\\
&&
Y_-(\la|\mu,\nu):=r(\la|\mu,\nu)\cdot Y_+(\la|\mu,\nu)
\label{Ypm}
\ena
The situation with positive integers $j>0$ is more complicated
since all three terms in the r.h.s. of (\ref{f2pm})
contribute to the residues at $\la=-1+3j$. We will not discuss it here.


\section{Construction of the matrix $G$}

As was pointed out above, we have set the fourth column of the matrix $G$
using the formula (\ref{gi}):
\bea
&&
g_{1,4}=(\la-\la^{-1})(\mu-\mu^{-1})(\nu-\nu^{-1})\Delta\bigl(f^{(0)}\bigr)(\la,\mu,\nu)
\nn\\
&&
g_{2,4}=(\la-\la^{-1})(\mu-\mu^{-1})(\nu-\nu^{-1})\Delta\bigl(f^{(1)}\bigr)(\la,\mu,\nu)
\nn\\
&&
g_{3,4}=(\la-\la^{-1})(\mu-\mu^{-1})(\nu-\nu^{-1})\Delta\bigl(f^{(2)}_+\bigr)
(\la,\mu,\nu)
\nn\\
&&
g_{4,4}=(\la-\la^{-1})(\mu-\mu^{-1})(\nu-\nu^{-1})\Delta\bigl(f^{(2)}_-\bigr)(\la,\mu,\nu)
\label{gi44}
\ena
The other columns of the matrix $G$ can be obtained as a linear combination of
the above fourth column without and with shifts of parameters $\la\rightarrow\la+3j,\;\mu\rightarrow\mu+3j,\;\nu\rightarrow\nu+3j$
with $j=-2,-1,0,1$. Using the transformation properties (\ref{transformf0},\ref{transformf1},\ref{transformf2}) of the
functions $f^{(0)},f^{(1)},f^{(2)}_{\pm}$, we can come to the following result:
\bea
G(\la,\mu,\nu)=
G_0(\la,\mu,\nu)+G_1(\la,\mu,\nu)+G_2(\la,\mu,\nu)\label{resG}
\ena
where
\bea
G_0(\la,\mu,\nu)=
\begin{pmatrix}
f^{(0)}(\la,\mu,\nu)\\
f^{(1)}(\la,\mu,\nu)\\
f_+^{(2)}(\la,\mu,\nu)\\
f_-^{(2)}(\la,\mu,\nu)
\end{pmatrix}\otimes
\begin{pmatrix}
v^{(0)}_1(\la,\mu,\nu),&v^{(0)}_2(\la,\mu,\nu),&v^{(0)}_3(\la,\mu,\nu),&
v^{(0)}_4(\la,\mu,\nu)
\end{pmatrix}
\label{G0}
\ena
\bea
G_1(\la,\mu,\nu)&=
&
G_{\psi}(\la|\mu,\nu)\cdot \psi\Bigl(\frac{1+\la}3\Bigr)\;
+\;G_{\psi}(\mu|\la,\nu)\cdot \psi\Bigl(\frac{1+\mu}3\Bigr)\;+\;
G_{\psi}(\nu|\la,\mu)\cdot \psi\Bigl(\frac{1+\nu}3\Bigr)\nn\\
&+&
G'_{\psi}(\la|\mu,\nu)\cdot \psi\Bigl(\frac{2+\la}3\Bigr)\;
+\;G'_{\psi}(\mu|\la,\nu)\cdot \psi\Bigl(\frac{2+\mu}3\Bigr)\;+
\;G'_{\psi}(\nu|\la,\mu)\cdot \psi\Bigl(\frac{2+\nu}3\Bigr)
\nn
\ena
with
\bea
&&
G_{\psi}(\la|\mu,\nu)=
\begin{pmatrix}
	0\\
	0\\
	1\\
	r(\la|\mu,\nu)
\end{pmatrix}\otimes
\begin{pmatrix}
	v^{(1)}_1(\la|\mu,\nu),&v^{(1)}_2(\la|\mu,\nu),&v^{(1)}_3(\la|\mu,\nu),&
	v^{(1)}_4(\la|\mu,\nu)
\end{pmatrix}
\label{Gpsi}
\ena
\bea
&&
G'_{\psi}(\la|\mu,\nu)\nn\\
&&=
\begin{pmatrix}
	0\\
	0\\
	1\\
	r(\la|\mu,\nu)
\end{pmatrix}\otimes
\begin{pmatrix}
	v'^{(1)}_1(-\la|-\mu,-\nu),
	&v^{(1)}_2(-\la|-\mu,-\nu),&v^{(1)}_3(-\la|-\mu,-\nu),&
	v^{(1)}_4(-\la|-\mu,-\nu)
\end{pmatrix},\nn\\
&&
\label{G'psi}\\
&&
v'^{(1)}_1(\la|\mu,\nu)=v^{(1)}_1(\la|\mu,\nu)
-2 x_0(\la,\mu,\nu)
\cdot v^{(1)}_4(\la|\mu,\nu),\nn\\
&&\nn\\
&&
x_0(\la,\mu,\nu):=\frac19(\la-\mu-\nu)(\mu-\la-\nu)(\nu-\la-\mu)
\label{x0}
\ena
\bea
G_2(\la,\mu,\nu)=
\begin{pmatrix}
0 & 0 & 0 & 0\\
v^{(2)}_{2,1} &v^{(2)}_{2,2} &v^{(2)}_{2,3} & v^{(2)}_{2,4}\\
v^{(2)}_{3,1} &v^{(2)}_{3,2} &v^{(2)}_{3,3} & 0\\
v^{(2)}_{4,1} &v^{(2)}_{4,2} &v^{(2)}_{4,3} & 0
\end{pmatrix}
\label{G2}
\ena
The explicit expressions for some of the functions $v^{(0)},v^{(1)},v^{(2)}$
look too bulky. Hence, it is hardly possible to explicitly present them.
In order to show explicit formulas we can represent the matrix $G$ in some other form:
\bea
G(\la,\mu,\nu)=\tilde{G}_0(\la,\mu,\nu)\;+\;\tilde{G}_1(\la,\mu,\nu)\;+\;
\tilde{G}_2(\la,\mu,\nu)
\label{Ga}
\ena
where the matrices $\tilde{G}_0,\tilde{G}_1,\tilde{G}_2$ can be
explicitly described:
\bea
\tilde{G}_0(\la,\mu,\nu)=
-\begin{pmatrix}
	f^{(0)}(\la,\mu,\nu)\\
	f^{(1)}(\la,\mu,\nu)\\
	f_+^{(2)}(\la,\mu,\nu)\\
	f_-^{(2)}(\la,\mu,\nu)
\end{pmatrix}\otimes
\begin{pmatrix}
	\overleftarrow{V_1}(\la,\mu,\nu),&\overleftarrow{V_2}(\la,\mu,\nu),&
	\overleftarrow{V_3}(\la,\mu,\nu),&
	\overleftarrow{V_4}(\la,\mu,\nu)
\end{pmatrix}
\label{G0a}
\ena
where the operators $\overleftarrow{V_i}$ act to the left on some function $f$
as follows:
\bea
&&
f(\la,\mu,\nu)\Bigl(\overleftarrow{V_i}(\la,\mu,\nu)\Bigr):=
f(\la,\mu,\nu)\Bigl(x_i^{(--)}(\la,\mu,\nu)\;\overleftarrow{\Delta^2_-}-
\overleftarrow{\Delta^2_-}\; x_0(\la,\mu,\nu)\delta_{i,1}+
x_i^{(+-)}(\la,\mu,\nu)\overleftarrow{\Delta_+}\;\overleftarrow{\Delta_-}\nn\\
&&
+
x_i^{(0)}(\la,\mu,\nu)\Bigr)(\la-\la^{-1})(\mu-\mu^{-1})(\nu-\nu^{-1})
\label{V_i}
\ena
and
\bea
x_1^{(--)}(\la,\mu,\nu)&=&
\frac1{24}\Bigl(\la^4+\mu^4+\nu^4\Bigr)
-\frac5{24}\Bigl(\la^3(\mu+\nu)+\mu^3(\la+\nu)+\nu^3(\la+\mu)\Bigr)\nn\\
&
+&\frac{185}{324}\Bigl(\la^2\mu^2+\la^2\nu^2+\mu^2\nu^2\Bigr)
+\frac{1913}{324}\Bigl(\la^2+\mu^2+\nu^2\Bigr)\nn\\
&
+&\frac{791}{108}\Bigl(\la\mu+\la\nu+\mu\nu\Bigr)
+\frac{13289}{216}
\label{x1--}\\
\nn\\
x_1^{(+-)}(\la,\mu,\nu)&=&
-\frac{175}{81}\Bigl(\la^2(\mu+\nu)+\mu^2(\la+\nu)+\nu^2(\la+\mu)\Bigr)
+
\frac{475}{216}\;\la\mu\nu-\frac{3983}{162}\Bigl(\la+\mu+\nu\Bigr)
\nn\\
\label{x1+-}\\
\nn\\
x_1^{(0)}(\la,\mu,\nu)&=&
\frac{55}6\Bigl(\la^2+\mu^2+\nu^2\Bigr)+\frac{473}{12}
\label{x10}\\
\nn\\
x_2^{(--)}(\la,\mu,\nu)&=&
5\Bigl(\la^2+\mu^2+\nu^2\Bigr)-\frac{146}{27}\Bigl(\la\mu+\la\nu+\mu\nu\Bigr)
+\frac{499}{27}
\label{x2--}\\
\nn\\
x_2^{(+-)}(\la,\mu,\nu)&=&-\frac23\,\Bigl(\la+\mu+\nu\Bigr)
\label{x2+-}\\
\nn\\
x_2^{(0)}(\la,\mu,\nu)&=&90
\label{x20}\\
\nn\\
x_3^{(--)}(\la,\mu,\nu)&=&
\frac1{12}\Bigl(\la^2+\mu^2+\nu^2\Bigr)-
\frac{7}{54}\Bigl(\la\mu+\la\nu+\mu\nu\Bigr)
-\frac{95}{108}
\label{x3--}\\
\nn\\
x_3^{(+-)}(\la,\mu,\nu)&=&\frac16\,\Bigl(\la+\mu+\nu\Bigr)
\label{x3+-}\\
\nn\\
x_3^{(0)}(\la,\mu,\nu)&=&\frac32
\label{x30}\\
\nn\\
x_4^{(--)}(\la,\mu,\nu)&=& -1,\quad
x_4^{(+-)}(\la,\mu,\nu)\;=\;
x_4^{(0)}(\la,\mu,\nu) \;=\;0
\label{x0all}
\ena
The operators $\Delta_{\pm}$ defined in (\ref{Deltapm})
act both on the coefficients $x_i^{- -},x_i^{+ -}$ and the function $f$.

Further we have
\bea
&&
\tilde{G}_1(\la,\mu,\nu)=
-\tilde{G}_{\psi}(\la|\mu,\nu)\cdot \psi\Bigl(\frac{1+\la}3\Bigr)
-\tilde{G}_{\psi}(\mu|\la,\nu)\cdot \psi\Bigl(\frac{1+\mu}3\Bigr)-
\tilde{G}_{\psi}(\nu|\la,\mu)\cdot \psi\Bigl(\frac{1+\nu}3\Bigr)\nn\\
&&+\;
\tilde{G}_{\psi}(-\la|-\mu,-\nu)\cdot \psi\Bigl(\frac{2+\la}3\Bigr)
+\tilde{G}_{\psi}(-\mu|-\la,-\nu)\cdot \psi\Bigl(\frac{2+\mu}3\Bigr)+
\tilde{G}_{\psi}(-\nu|-\la,-\mu)\cdot \psi\Bigl(\frac{2+\nu}3\Bigr)
\nn\\
\label{tildeG1}
\ena
with
\bea
&&
\tilde{G}_{\psi}(\la|\mu,\nu)=
\label{tildeGpsi}
\\
&&
{\ds\begin{pmatrix}
	0\\
	0\\
	1\\
	r(\la|\mu,\nu)
\end{pmatrix}\otimes
\begin{pmatrix}
	a_1(\la|\mu,\nu)c_1(\la|\mu,\nu)+a_2(\la|\mu,\nu)c_2(\la|\mu,\nu),
	&\frac{28}{27}\,c_1(\la|\mu,\nu),&\frac{1}{27}\,c_1(\la|\mu,\nu),&
	0
\end{pmatrix}
}
\nn\\
\label{tildeGpsi1}\\
&&
a_1(\la|\mu,\nu)=\frac{1}{54}\;\Bigl(\la^2+\mu^2+\nu^2\Bigr)
-\frac{767}{1944}\;\la\Bigl(\mu+\nu\Bigr)+\frac{427}{1944}\;\mu\nu
-\frac{115}{216}\;\la+\frac{1141}{1296}\;\Bigl(\mu+\nu\Bigr)-\frac{55}{3888}
\nn
\\
&&
a_2(\la|\mu,\nu)=-\frac{73}{162}\la+\frac{743}{648}\Bigl(\mu+\nu\Bigr)+\frac{199}{108}
\label{a1a2}
\ena
and finally the matrix $\tilde{G}_2$ looks:
\bea
\tilde{G}_2(\la,\mu,\nu)=
\begin{pmatrix}
	0 & 0 & 0 & 0\\
	d_{2,1}(\la,\mu,\nu) &0 &0 & 0\\
	d_{3,1}(\la,\mu,\nu)&d_{3,2}(\la,\mu,\nu) &d_{3,3}(\la,\mu,\nu) & 0\\
	d_{4,1}(\la,\mu,\nu) &d_{4,2}(\la,\mu,\nu) &d_{4,3}(\la,\mu,\nu) & 0
\end{pmatrix}
\label{tildeG2}
\ena
with
\bea
&&
d_{2,1}(\la,\mu,\nu)=\frac{53}{36}\nn\\
&&
d_{3,1}(\la,\mu,\nu)=
\frac{t'(\la|\mu,\nu)-t'(\mu|\la,\nu)}{r(\la|\mu,\nu)-r(\mu|\la,\nu)}
\nn\\
&&
d_{3,2}(\la,\mu,\nu)=-\frac{64}9 d(\la,\mu,\nu),\quad
d_{3,3}(\la,\mu,\nu)=\frac29\; d(\la,\mu,\nu),\nn\\
&&
d(\la,\mu,\nu)=
\frac{t(\la|\mu,\nu)-t(\mu|\la,\nu)}{r(\la|\mu,\nu)-r(\mu|\la,\nu)}\nn\\
&&
d_{4,1}(\la,\mu,\nu)=
=\frac{t'(\la|\mu,\nu)r(\mu|\la,\nu)-t'(\mu|\la,\nu)r(\la|\mu,\nu)}{r(\la|\mu,\nu)-r(\mu|\la,\nu)}\nn\\
&&
d_{4,2}(\la,\mu,\nu)=-\frac{64}9\; d'(\la,\mu,\nu),\quad d_{4,3}(\la,\mu,\nu)=\frac29 d'(\la,\mu,\nu),\nn\\
&&
d'(\la,\mu,\nu)=\frac{t(\la|\mu,\nu)r(\mu|\la,\nu)-t(\mu|\la,\nu)r(\la|\mu,\nu)}{r(\la|\mu,\nu)-r(\mu|\la,\nu)}
\label{d}\\
\nn\\
&&
t(\la|\mu,\nu)=\frac{\la^2-1}{\mu\nu(2\la-\mu-\nu)}\cdot 
\Bigl(\mu(\la-\nu)+\nu(\la-\mu)-2\Bigr)
\label{t}\\
&&
t'(\la|\mu,\nu)=\frac{\la^2-1}{\mu\nu(2\la-\mu-\nu)}\cdot
\Bigl(\frac19\,\la^3(\mu+\nu)+\frac{70}{27}\,\la^2(\mu^2+\nu^2)+
\frac{26}{27}\,\la^2\mu\nu+\frac19\,\la(\mu^3+\nu^3)\label{t'}
\\
&&
-\frac{175}{27}
\la\mu\nu(\mu+\nu)+\frac{184}{27}\mu^2\nu^2-\frac29\mu\nu(\mu^2+\nu^2)
-\frac{38}9(\la^2+\mu^2+\nu^2)-\frac{152}{27}\la(\mu+\nu)+\frac{532}{27}\mu\nu
+\frac{188}9\Bigr)\nn\\
\nn
\ena
The above formulae for the elements $d_{i,j}$ look non-symmetric but
one can check that they actually are symmetric wrt all their arguments
$\la,\mu,\nu$.

\noindent
{\bf Remark 9.}
It is interesting to look at the determinant of the matrix $G$
which fulfills the one-dimensional difference equation:
\bea
\det{G(\la,\mu,\nu)} \cdot \det{A(\la,\mu,\nu)} =
\det{G(\la-3,\mu-3,\nu-3)}
\label{detGeq}
\ena
This equation can be easily solved because we have a simple formula
for determinant of the matrix $A$ easily deduced from (\ref{detM}):
\bea
\det{A(\la,\mu,\nu)} =\frac{\la^2_{(-4)}\;\la^3_{(-2)}\;\la^3}{\la^3_{(-3)}\;
	\la^3_{(-3)}\;\la^3_{(-1)}
\la^2_{(1)}}\cdot
\frac{\mu^2_{(-4)}\;\mu^3_{(-2)}\;\mu^3}{\mu^3_{(-3)}\;
	\mu^3_{(-3)}\;\mu^3_{(-1)}\;
	\mu^2_{(1)}}\cdot
\frac{\nu^2_{(-4)}\;\nu^3_{(-2)}\;\nu^3}{\nu^3_{(-3)}\;\nu^3_{(-3)}\;
	\nu^3_{(-1)}\;
	\nu^2_{(1)}}
\label{detA}
\ena
The solution looks
\bea
\det{G(\la,\mu,\nu)}=\phi(\la,\mu,\nu)\cdot
\frac{(\la^2-1)^2}{\la^3}\cdot
\frac{(\mu^2-1)^2}{\mu^3}\cdot
\frac{(\nu^2-1)^2}{\nu^3}\cdot
f^{(0)}(\la,\mu,\nu)
\label{detG}
\ena
with a quasi-constant $\phi$. It is interesting that it
depends only on the function $f^{(0)}$ and not on the other
functions $f^{(1)}, f^{(2)}_{\pm}$
that present in solution (\ref{resG}) or (\ref{Ga}).
It can be seen by taking the determinant of (\ref{resG})
even without specifying the functions $v^{(0)}_i,v^{(1)}_i,
v^{(2)}_{i,j}$. We can check that after substituting explicit
expressions for these functions, we come to the above result
(\ref{detG}) with the following quasi-constant:
\bea
\phi(\la,\mu,\nu)=\frac{p_2-3}{27}
\label{phi}
\ena
with the polynomial $p_2$ given by (\ref{p2}).

\section{Inverse of $G$}

The formulae of the previos section which define the matrix $G$ look a bit
complicated. It turns out that the inverse matrix $G^{-1}$ looks simpler.
Since it may appear to be more convenient in future to use  $G^{-1}$ istead of $G$,
we decided to explicitly show $G^{-1}$ also.
We will call it
$$
G^{(i)}(\la,\mu,\nu):= G^{-1}(\la,\mu,\nu)
$$
Obviously, it fulfills the equation
\bea
A(\la,\mu,\nu)G^{(i)}(\la-3,\mu-3,\nu-3) = G^{(i)}(\la,\mu,\nu)
\label{Gi}
\ena
As appeared, the matrix $G^{(i)}$ has sort of a ``nested'' structure: the first
column is defined through three other columns, the second column can be
defined in terms of the third and fourth columns. Let $g^{(i)}_{j,k}$ be the
elements of the matrix $G^{(i)}$. Then
\bea
&&G^{(i)}=\label{Gidef}\\
&&
{\ds
\begin{pmatrix}
g^{(i)}_{1,2}\\
g^{(i)}_{2,2}\\
g^{(i)}_{3,2}\\
g^{(i)}_{4,2}
\end{pmatrix}
\otimes
\begin{pmatrix}
-\frac{f^{(1)}}{f^{(0)}},&1,&0,&0
\end{pmatrix}
+
\begin{pmatrix}
	g^{(i)}_{1,3}\\
	g^{(i)}_{2,3}\\
	g^{(i)}_{3,3}\\
	g^{(i)}_{4,3}
\end{pmatrix}
\otimes
\begin{pmatrix}
	-\frac{f^{(2)}_+}{f^{(0)}},&0,&1,&0
\end{pmatrix}
+
\begin{pmatrix}
	g^{(i)}_{1,4}\\
	g^{(i)}_{2,4}\\
	g^{(i)}_{3,4}\\
	g^{(i)}_{4,4}
\end{pmatrix}
\otimes
\begin{pmatrix}
	-\frac{f^{(2)}_-}{f^{(0)}},&0,&0,&1
\end{pmatrix}
}
\nn\\
&&\quad \quad
+\begin{pmatrix}
	z_{1}\\
	z_{2}\\
	z_{3}\\
	z_{4}
\end{pmatrix}
\otimes
\begin{pmatrix}
	\frac{1}{f^{(0)}},&0,&0,&0
\end{pmatrix}
\nn
\ena
where
\bea
z_{j}=w^{(0)}_j(\la,\mu,\nu) &+&
\Bigl(\chi_1(\la|\mu,\nu)\;\psi\bigl(\frac{1+\la}3\bigr)\;+\;
\chi_1(-\la|-\mu,-\nu)\;\psi\bigl(\frac{2+\la}3\bigr)\Bigr)\cdot w_j(\la|\mu,\nu)\nn\\
&+&
\Bigl(\chi_1(\mu|\la,\nu)\;\psi\bigl(\frac{1+\mu}3\bigr)\;+\;
\chi_1(-\mu|-\la,-\nu)\;\psi\bigl(\frac{2+\mu}3\bigr)\Bigr)\cdot w_j(\mu|\la,\nu)
\label{z}\\
&+&
\Bigl(\chi_1(\nu|\la,\mu)\;\psi\bigl(\frac{1+\nu}3\bigr)\;+\;
\chi_1(-\nu|-\la,-\mu)\;\psi\bigl(\frac{2+\nu}3\bigr)\Bigr)\cdot w_j(\nu|\la,\mu)
\nn
\ena
\bea
&&\chi_1(\la|\mu,\nu)=-\frac{\frac{\la-1}2\bigl((\mu-\nu)^2-1\bigr)
	+\frac{\la+1}6 \Bigl(2\la-\mu-\nu-3\Bigr)^2}{(\la-\mu-1)(\la-\nu-1)\bigl((\mu-\nu)^2-1\bigr)}
\label{chi}\\
\nn\\
&&
w^{(0)}_j(\la,\mu,\nu)=\frac{\la\mu\nu}{p_2-3}\cdot
\Biggl(\frac{v_j^+}{(\la+1)(\mu+1)(\nu+1)}-\frac{v_j^-}{(\la-1)(\mu-1)(\nu-1)}\Biggr)
\label{w0}\\
\nn\\
&&
v_1^{\pm}=0,\quad v_2^{\pm}=-\frac1{80},\quad v_3^{\pm}=-\frac25,\quad
v_4^{\pm}=-\frac{23}{240}\,s_1^2+\frac{14}{45}\,s_2\pm \frac{s_1}{12}+\frac7{30}
\label{vpm}\\
\nn\\
&&
w_j(\la|\mu,\nu)=\chi_2(\la|\mu,\nu)\cdot \Bigl(y_2(\la+1|\mu+1,\nu+1)u^+_j-
y_2(\la-1|\mu-1,\nu-1)u^-_j + d^{(i)}_j\Bigr)
\label{wj}\\
\nn\\
&&
\chi_2(\la|\mu,\nu)= \frac{(\mu-\nu)^2-1}{(\la-\mu)(\la-\nu)(p_2-3)},\quad
y_2(\la|\mu,\nu)=\frac{\la^2-1}{48\mu\nu}
\label{y2}\\
&&
d^{(i)}_j = -\frac{1}6\;\la \cdot e^{(i)}_j\;+\; \frac7{90}s_3\cdot \delta_{j,4},\quad
e^{(i)}_1=0,\quad
e^{(i)}_2=\frac9{40},\quad e^{(i)}_3=-\frac{63}{10},\quad
e^{(i)}_4=\frac{101}{30}+\frac{3}{5} s_1^2-\frac{17}{15}s_2
\label{ei}\\
\nn\\
&&
u_1^{\pm}=1,\quad u_2^{\pm}=\frac15 s_2\pm \frac3{10}s_1+\frac1{10},
\quad
u_3^{\pm}= -\frac{s_1^2}2-\frac{64}{15}s_2\mp \frac{42}{5}s_1+\frac{116}{5}
\label{upm}\\
\nn\\
&&
u_4^+\equiv u_4^+(\la,\mu,\nu)
=-\frac{32}5+\frac{202}{45}s_1+\frac{61}{18}s_1^2+\frac{31}{45}s_1^3-
\frac{244}{45}s_2
-\frac{16}{15}s_1s_2+\frac8{15}s_1^2s_2+\frac{16}9s_2^2
-\frac{124}{45} s_3-\frac{32}{45}s_1 s_3\nn\\
\nn\\
&&
u_4^-\equiv u_4^-(\la,\mu,\nu)=u_4^+(-\la,-\mu,-\nu)-\frac29(\la-\mu-\nu)(\mu-\la-\nu)(\nu-\la-\mu)
\nn
\ena
For the second column we got
\bea
&&
g^{(i)}_{j,2}=
w_j(\la|\mu,\nu)\Bigl(\psi\bigl(\frac{1+\la}3\bigr)+
\psi\bigl(\frac{2+\la}3\bigr)\Bigr)+
w_j(\mu|\la,\nu)\Bigl(\psi\bigl(\frac{1+\mu}3\bigr)
+
\psi\bigl(\frac{2+\mu}3\bigr)\Bigr)
\label{gi2}\\
\nn\\
&&
+w_j(\nu|\la,\mu)\Bigl(\psi\bigl(\frac{1+\nu}3\bigr)+
\psi\bigl(\frac{2+\nu}3\bigr)\Bigr)
+\frac{\la\mu\nu}{p_2-3} \Bigl(\gamma_3v_j^+\;+\;\gamma_{-3}v_j^-\;+\;v^0_j\Bigr),
\nn\\
\nn\\
&&
v^0_j = -v^+_j - v^-_j -\frac32\cdot \delta_{j,4}
\nn
\ena
For the third column we have:
\bea
&&
g^{(i)}_{j,3}=y_3(\la-1,\mu-1,\nu-1)\;u^-_j\;-\;y_3(\la+1,\mu+1,\nu+1)\;
u^+_j
\;+\; e^{(i)}_j\label{gi3}\\
\nn\\
&&
y_3(\la,\mu,\nu)=\frac1{\la\mu\nu}\cdot\Bigl(\frac{s_1^2}{12}-\frac{s_2}8 -\frac{1}8\Bigr)
\label{y3}
\ena
where $e^{(i)}_j$ were defined above in (\ref{ei}) and
finally:
\bea
&&
g^{(i)}_{j,4}=y_4(\la-1,\mu-1,\nu-1)\;u^-_j\;-\;y_4(\la+1,\mu+1,\nu+1)\;
u^+_j
\label{gi4}\\
\nn\\
&&
y_4(\la,\mu,\nu)=\frac{s_1}{24\la\mu\nu}
\label{y4}
\ena
Above we mentioned that $G^{(i)}$ has a kind of ``nested'' structure. This seems
to be similar concerning the transcendental properties also: the first column,
being combination of the functions\\ $1/f^{(0)}, f^{(1)}/f^{(0)}, f_{\pm}^{(2)}/f^{(0)}$
with coefficients which depend on logarithmic derivative of the $\Gamma$-function
$\psi$, has the most complicated transcendental properties. The second column
depends only on the $\psi$-function with rational coefficients.
The most unexpected fact for us is that the third and the fourth column
of the inverse matrix $G^{(i)}=G^{-1}$ are pure rational. It looks a bit strange because all elements of the matrix $G$ itself are transcendental.

\noindent
{\bf Remark 10.}
Actually, the rational vector $w(\la|\mu,\nu)$ with entries $w_j(\la|\mu,\nu),\;
j=1,\cdots,4$ which are the coefficients standing with
$\psi$-functions in $g^{(i)}_{2,j}$
defined above in (\ref{wj}) fulfills itself the difference equation
(\ref{Gi}) written for the column:
\bea
A(\la,\mu,\nu) w(\la-3|\mu-3,\nu-3)=w(\la|\mu,\nu)
\label{w}
\ena
This is a rational solution which is not symmetric wrt
transpositions $\la\leftrightarrow \mu,
\la\leftrightarrow \nu$. This solution turned out to be linear dependent
on the above third and fourth columns (\ref{gi3},\ref{gi4})
which are symmetric wrt all its arguments $\la,\mu,\nu$
\bea
&&
w_j(\la|\mu,\nu)= -\frac1{18}\;(2\la-\mu-\nu)\; \chi_2(\la|\mu,\nu)\cdot \Bigl( g^{(i)}_{j,3}(\la,\nu,\mu)+
r(\la,\mu,\nu)\;g^{(i)}_{j,4}(\la,\nu,\mu)\Bigr)
\label{wg3g4}
\ena
\noindent
{\bf Remark 11.}
Having in mind the formula (\ref{formA}) for the matrix $A(\la,\mu,\nu)$, it is interesting to see how the relation (\ref{Gi}) works when we approach the poles of the matrix $A$, for example,
the pole $\la=3$. As we know the residue $M^{(3)}_2$ has the rank 3. Since
the r.h.s. of (\ref{Gi}) does not have poles at the point $\la=3$, every column of the
matrix $G^{(i)}(0,\mu-3,\nu-3)$ must belong to the kernel of $M^{(3)}_2$:
\bea
M^{(3)}_2 G^{(i)}(0,\mu-3,\nu-3) = 0
\label{kerGi}
\ena
which means that the matrix $G^{(i)}(0,\mu-3,\nu-3)$ must have rank 1.
Indeed, we can check that every column of $G^{(i)}(0,\mu-3,\nu-3)$ is proportional
to the null vector of the matrix $M^{(3)}_2$.

\noindent
{\bf Remark 12.}
Let us make one more remark about the pole structure of the matrix $G^{(i)}$. In subsection 4.3 we made a remark 3. where we pointed
out that since the difference between $\psi(\frac{2+\la}3)$
and $\psi(\frac{1-\la}3)$ is a quasi-constant, we can freely choose
in our solutions for $h^{(1)}_{\pm}$ (see (\ref{h1+})) one of these two possiblities.
The same argument works for the choice of either $\psi(\frac{2-\la}3)$
or $\psi(\frac{1+\la}3)$. We took the variant in which the functions $f^{(2)}_{\pm}$ do not have poles of second order.
The drawback of this choice is that some of the elements of the  above inverse matrix $G^{(i)}$ have poles of second order. Using the same
logic we used in the remark 3, we can avoid such poles if we choose
in the rhs of (\ref{h1+}) the function $\psi(\frac{2-\la}3)$ instead of $\psi(\frac{1+\la}3)$.
In this way we can define the matrix $\tilde{G}^{(i)}$ in the same
way as in (\ref{Gidef}) where
we have to substitute everywhere $\psi(\frac{1+\la}3)$ by
$\psi(\frac{2-\la}3)$ (and the same for $\psi$ of arguments with
$\mu$ and $\nu$)
and take the functions ${\tilde{f}}^{(2)}_{\pm}$
instead of $f^{(2)}_{\pm}$ with the transform:
\bea
{\tilde{f}}_{\pm}^{(2)}(\la+3,\mu+3,\nu+3)=
\gamma_3(\la,\mu,\nu)\;{\tilde{f}}_{\pm}^{(2)}(\la,\mu,\nu)-
\beta_3(\la,\mu,\nu){\tilde{h}}_{\pm}^{(1)}(\la,\mu,\nu)
\label{tildetransformf2}
\ena
where
\bea
{\tilde{h}}_{\pm}^{(1)}(\la,\mu,\nu)=
&
b_{1,\pm}(\la|\mu,\nu)\cdot\psi(\frac{2-\la}3)\;+
\;b_{2,\pm}(\la|\mu,\nu)\cdot\psi(\frac{2+\la}3)\nn\\
+&
b_{1,\pm}(\mu|\la,\nu)\cdot\psi(\frac{2-\mu}3)\;+
\;b_{2,\pm}(\mu|\la,\nu)\cdot\psi(\frac{2+\mu}3)\nn\\
+&
b_{1,\pm}(\nu|\la,\mu)\cdot \psi(\frac{2-\nu}3)\;+\;b_{2,\pm}(\nu|\la,\mu)\cdot\psi(\frac{2+\nu}3)
\label{tildeh1+}
\ena
with the same functions $b_{1,\pm},b_{2,\pm}$ defined in (\ref{b1b2+-}).

In fact, we can check that this substitutions correspond to
the adding some linear combinations of the third and the fourth
columns of $G^{(i)}$ with quasi-constant coefficients
to the first and the second columns. It is a bit more tricky
to see how it happens for the first column. In fact, it is based
on the following identity:
\bea
&&
\Bigl(-\frac{1}{\la_{(-1)}\mu_{(-1)}\nu_{(-1)}}+\chi_1(\la-3|\mu-3,\nu-3)\Bigr) w_j(\la-3|\mu-3,\nu-3)\nn\\
&&
-\frac{b_{1,+}(\la-3|\mu-3,\nu-3)}{\la_{(-1)}\mu_{(-1)}\nu_{(-1)}}
\Bigl(g^{(i)}_{j,3}(\la-3,\mu-3,\nu-3)+r(\la|\mu,\nu)
g^{(i)}_{j,4}(\la-3,\mu-3,\nu-3)\Bigr)\nn\\
&&
=\gamma_{-3}(\la,\mu,\nu)\, \chi_1(\la|\mu,\nu)\;w_j(\la-3|\mu-3,\nu-3)
\label{ident}
\ena
which can be shown using the relation (\ref{wg3g4}).

It is obwious that the above described modified matrix ${\tilde{G}}^{(i)}$ will satisfy the equation (\ref{Gi}) also.

\section{Acknowledgments}
The authors would like to thank F. G{\"o}hmann and A. Kl{\"u}mper
for stimulating discussions.
HB acknowledges financial support by the DFG in the framework of the
research unit FOR 2316. API is grateful to the Heisenberg-Landau program for financial support and the Physics Department, University of Wuppertal for kind hospitality during his visit in early 2021.

\section{Appendix A}

	The coefficient $S_6$ is represented in (\ref{S6}),
 where the homogeneous symmetric polynomials $\bar{S}_6^{(k)}$ can be explicitly expressed
 in terms of the basic symmetric polynomial (\ref{sypol})
 {\footnotesize
\begin{equation}
\label{s6bar}
\begin{array}{c}
\bar{S}_6^{(16)} = 18 s_3^4 (s_2^2-3 s_3 s_1) \; , \;\;\;
\bar{S}_6^{(15)} = 6 s_3^3
(14 s_2^3 - 45 s_3s_2s_1 + 27 s_3^2) \; , \\ [0.2cm]
\bar{S}_6^{(14)} = - 3 s_3^2 (38 s_2^4 -322 s_3 s_2^2 s_1
-555 s_3^2 s_2 +813 s_3^2 s_1^2) \; ,
\\ [0.2cm]
\bar{S}_6^{(13)} = -2 s_3 (490 s_2^5-2653 s_3 s_2^3 s_1
-2082 s_3^2 s_2^2+4257 s_3^2 s_2 s_1^2+90 s_3^3 s_1) \; ,
\\ [0.2cm]
\bar{S}_6^{(12)} = -280 s_2^6-3824 s_3 s_2^4 s_1
-12993 s_3^2 s_2^3 +28713 s_3^2 s_2^2 s_1^2 + \\ [0.1cm]
+35286 s_3^3 s_2 s_1
-44262 s_3^3 s_1^3 +33525 s_3^4 \; ,
\\ [0.2cm]
\bar{S}_6^{(11)} =
- 20 s_2^5 s_1 + s_3  (20682 s_2^3 s_1^2 -86238  s_2^4 )
+ \\ [0.1cm]
+ s_3^2( 252816  s_2^2 s_1 - 44850  s_2 s_1^3)
+s_3^3 (118266  s_2 -169758  s_1^2) \; , \\ [0.2cm]
\bar{S}_6^{(10)} = -20428 s_2^5 +14132 s_2^4 s_1^2
+s_3 (62226 s_2^2 s_1^3  -190304 s_2^3 s_1)
+  \\ [0.1cm]
+ s_3^2 (-572829 s_2^2
+776319 s_2 s_1^2-279351 s_1^4) +701598 s_3^3 s_1 \; , \\ [0.2cm]
\bar{S}_6^{(9)} =
7920 s_2^4 s_1 +18820 s_2^3 s_1^3
+s_3 (1589178 s_2^2 s_1^2 -2715150 s_2^3
-233574 s_2 s_1^4) + \\ [0.1cm]
+ s_3^2 (2807946 s_2 s_1- 1544880s_1^3) +2157294 s_3^3 \; ,
\\ [0.2cm]
\bar{S}_6^{(8)} =
-696948 s_2^4 +479780 s_2^3 s_1^2 -127260 s_2^2 s_1^4
\\ [0.1cm]
+ s_3 (1948284 s_2 s_1^3 -848322 s_2^2 s_1
-457218 s_1^5)
+ s_3^2 (-13018131 s_2 +4564524 s_1^2) \; ,
\\ [0.2cm]
\bar{S}_6^{(7)} =-1120260 s_2^3 s_1
+64200 s_2^2 s_1^3 -131520 s_2 s_1^5 \\ [0.1cm]
+s_3 (-34921154 s_2^2
+ 28658244 s_2 s_1^2-4001424 s_1^4)+7333568 s_3^2 s_1  \; ,
\\ [0.2cm]
\bar{S}_6^{(6)} = -12994564 s_2^3
-298176 s_2^2 s_1^2 +2414016 s_2 s_1^4 - 353232 s_1^6 +
\\ [0.1cm]
+s_3 (35325316 s_2 s_1-4446324 s_1^3) - 97078341 s_3^2 \; ,
\end{array}
\end{equation}
\begin{equation}
\label{s6bar0}
\begin{array}{c}
\bar{S}_6^{(5)} = -47190920 s_2^2 s_1 +32696640 s_2 s_1^3
-3633600 s_1^5 +s_3 (4887360 s_1^2-128181870 s_2)) \; ,
\\ [0.2cm]
\bar{S}_6^{(4)} =  -128962084 s_2^2
+ 49506848 s_2 s_1^2 -2894112 s_1^4 - 38373002 s_3 s_1 \; ,
\\ [0.2cm]
\bar{S}_6^{(3)} =
-238864320 s_2 s_1 +51280000 s_1^3 + 191216592 s_3 \; ,
\\ [0.2cm]
\bar{S}_6^{(2)} = -345611808 s_2 + 66051632 s_1^2 \; , \;\;
\bar{S}_6^{(1)} = -324043200 s_1 \; , \;\;
\bar{S}_6^{(0)} = -830311488 \; .
\end{array}
\end{equation}
}

\vspace{0.5cm}

For coefficient $S_3$ we have expansion (\ref{s3})
where for homogeneous symmetric polynomials $\bar{S}_3^{(k)}$
with degree of homogeneity $k$ we have explicit formulas
{\footnotesize  \begin{equation}
	\label{coef3}
	\begin{array}{c}
	S_3^{(0)}  = -60234812928 \; , \;\;\;
	S_3^{(1)}  = -39883264128 s_1 \; , \;\;\;
	S_3^{(2)}  = -(30437149536 \, s_2+1346775808 \, s_1^2)\; ,  \\ [0.3cm]
	S_3^{(3)}  = -25740363168 s_2 s_1 +4983182592 s_1^3+11386810584 s_3 \; ,
	\\ [0.3cm]
	S_3^{(4)}  = -11688413032 s_2^2-688735040 s_2 s_1^2
	+775145728 s_1^4  + 1259128384 s_3 s_1 \; , \\ [0.3cm]
	S_3^{(5)}  = -7390237944 s_2^2 s_1 +3585464448 s_2 s_1^3-314201472 s_1^5
	- s_3 (8510548338 s_2 +143022168 s_1^2) \; ,
	\end{array}
	\end{equation}
	\begin{equation}
	\label{coef3d}
	\begin{array}{c}
	S_3^{(6)}  =  -1814990288 s_2^3 -670287888 s_2^2 s_1^2
	+815605344 s_2 s_1^4-97185792 s_1^6 + \\ [0.2cm]
	-s_3(429473998 s_2s_1 + 258924848 s_1^3)
	-5700786663 s_3^2\; , \\ [0.3cm]
	S_3^{(7)}  =  -656035104 s_2^3 s_1+198736464 s_2^2 s_1^3
	+21493536 s_2 s_1^5-7425792 s_1^7  + \\ [0.2cm]
	+s_3 (-3368317620 s_2^2+2809327932 s_2s_1^2-388095528 s_1^4)
	-1401020871 s_3^2 s_1 \; , \\ [0.3cm]
	S_3^{(8)}  =  -149006512 s_2^4+18985664 s_2^3 s_1^2
	+8009976 s_2^2 s_1^4 -5218752 s_2 s_1^6 + \\ [0.2cm]
	+s_3 (-684172968 s_2^2 s_1 +760575108 s_2 s_1^3-117049104 s_1^5)
	+s_3^2 (-1407865098 s_2 + 308714750 s_1^2)\; , \\ [0.3cm]
	S_3^{(9)}  = -25300368 s_2^4 s_1
	+14921760 s_2^3 s_1^3 -4107480 s_2^2 s_1^5 + \\ [0.2cm]
	+s_3 (-414864444 s_2^3 +250290288 s_2^2 s_1^2
	+12632790 s_2 s_1^4 -10477656 s_1^6) + \\ [0.2cm]
	+s_3^2 (40816704 s_2 s_1 -39235770 s_1^3)-17889732 s_3^3\; , \\ [0.3cm]
	S_3^{(10)}  = -7170976 s_2^5+4945072 s_2^4 s_1^2-581360 s_2^3 s_1^4
	+s_3(-79313012 s_2^3 s_1 +51441576 s_2^2 s_1^3
	-8454870 s_2 s_1^5) + \\ [0.2cm]
	+s_3^2 (-161625386 s_2^2+187592016 s_2 s_1^2
	-54973719 s_1^4)+ 66540743  s_3^3 s_1 \; , \\ [0.3cm]
	\end{array}
	\end{equation}
	\begin{equation}
	\label{coef3b}
	\begin{array}{c}
	S_3^{(11)}  =  -524928 s_2^5 s_1 +477072 s_2^4 s_1^3
	+s_3 (-23999832 s_2^4
	+8695116 s_2^3 s_1^2 -882876 s_2^2 s_1^4) + \\ [0.2cm]
	+s_3^2 (17392050 s_2^2 s_1 +12468240 s_2 s_1^3-7018047 s_1^5)
	+s_3^3 (-4071798 s_2 +5825883 s_1^2)\; , \\ [0.3cm]
	S_3^{(12)}  =  -201128 s_2^6 +113952 s_2^5 s_1^2
	+s_3(-3089512 s_2^4 s_1 +808900 s_2^3 s_1^3) + \\ [0.2cm]
	+s_3^2 (-10282200 s_2^3+15571278 s_2^2 s_1^2 -3351942s_2 s_1^4)
	+s_3^3 (14089119 s_2 s_1 -8620239 s_1^3)+ 3254610 s_3^4 \; , \\ [0.3cm]
	S_3^{(13)}  =  -8760 s_2^6 s_1
	+ s_3 (-678018 s_2^5+136368 s_2^4 s_1^2)
	+s_3^2 (1069896 s_2^3 s_1 +347346 s_2^2 s_1^3) + \\ [0.2cm]
	+ s_3^3 (-548718 s_2^2+1835556 s_2 s_1^2-1647279 s_1^4)
	+1264101 s_3^4 s_1 \; ,
	\\ [0.3cm]
	S_3^{(14)}  =  -2800 s_2^7 -38286 s_3 s_2^5 s_1
	+ s_3^2 (-327639 s_2^4+348332 s_2^3 s_1^2) + \\ [0.2cm]
	+s_3^3 (1017417 s_2^2 s_1-605001 s_2 s_1^3)
	+ s_3^4 (478626 s_2 -436356 s_1^2)   \; , \\ [0.3cm]
	S_3^{(15)}  =   -7932 s_3 s_2^6 + 27741 s_3^2 s_2^4 s_1
	+s_3^3 (-32970 s_2^3+ 36093 s_2^2 s_1^2) + \\ [0.2cm]
	+ s_3^4 (198354 s_2 s_1-179343 s_1^3)+36972 s_3^5 \; , \\ [0.3cm]
	S_3^{(16)}  = -3590  s_3^2 s_2^5  + 22721 s_3^3  s_2^3 s_1
	+ s_3^4(25566 s_2^2 -44568 s_2 s_1^2) - 423 s_3^5 s_1 \; , \\ [0.3cm]
	S_3^{(17)}  = -78 s_3^3 s_2^4  + 2733  s_3^4 s_2^2 s_1
	+ s_3^5 (7920 s_2-10197 s_1^2) \; ,
	\\ [0.3cm]
	S_3^{(18)}  =
	15 s_3^4 (27 s_3^2 + 26 s_2^3 - 81 s_3 s_2 s_1) \; , \;\;\;
	S_3^{(19)}  = 90 s_3^5 (s_2^2 - 3 s_3 s_1 )\; .
	\end{array}
	\end{equation}
}

\section{Appendix B}

In this appendix we define the elements of the matrix $A$
using the formula (\ref{formA}). Since the  residues
$M_i^{(a)}, a=3, 1, -1$ look much simpler, we will present them
here. Actually, it is enough to give formulae only for $M_2^{(a)}$
because the others can be restored using symmetry. In analog
with (\ref{sypol}) we will use symmetric polynomials of two variables:
\bea
s'_1 =  \mu + \nu,\quad
s'_2 =  \mu \nu
\label{sypola}
\ena
Introduce the matrices $M^{(a)}$ as symmetric matrix functions of $\mu,\nu$ that we write via $s'_1,s'_2$:
\bea
M^{(a)}(\mu,\nu)=\sum_{k=0}^5 ({s'_2})^k\; M^{(a,k)},\quad
a=3,1,-1
\label{Mamunu}
\ena
with the matrix coefficients $M^{(a,k)}$ which depend on one variable $s'_1$ and will be explicitly shown below.
Then for the residues $M_i^{(a)}$ in (\ref{formA}) we have:
\bea
&&
M_2^{(a)}={D^{(a)}}(\mu-\la+a,\nu-\la+a)\cdot M^{(a)}(\mu-\la+a,\nu-\la+a),
\nn\\
&&
M_3^{(a)}={D^{(a)}}(\la-\mu+a,\nu-\mu+a)\cdot M^{(a)}(\la-\mu+a,\nu-\mu+a),
\label{Ma}\\
&&
M_4^{(a)}={D^{(a)}}(\la-\nu+a,\mu-\nu+a)\cdot M^{(a)}(\la-\nu+a,\mu-\nu+a),
\nn\\
\ena
where the multiplier $D^{(a)}$ is defined through the denominator in (\ref{A}):
\bea
&&{\ds {D^{(a)}}(\mu,\nu):=\res_{\la=a}\Biggl(\frac{1}{D(\la,\mu,\nu)}\Biggr)
}\label{Da}\\
&&
D(\la,\mu,\nu)=(\la-3)(\la-1)(\la+1)(\mu-3)(\mu-1)(\mu+1)
(\nu-3)(\nu-1)(\nu+1)
\nn
\ena

\vspace{0.5cm}

The lower-triangular matrix (\ref{M-}) $M_-$  explicitly looks
\bea
M_-=
\begin{pmatrix}
0 & 0 & 0 & 0\\
\\
0 & 0 & 0 & 0\\
\\
m_{3,1} & 0 & 0 & 0\\
\\
m_{4,1} & m_{4,2} & m_{4,3} & 0
\end{pmatrix}
\label{Mma}
\ena
\bea
&&
m_{3,1}=\frac{1089}{64}-\frac{11}{12}\;s_1,\nn\\
\nn\\
&&
m_{4,1}=-\frac{1673}{576}\;-\;\frac{17}{36}\;s_1\;+\;
\frac{183}{16}\;s_1^2\;+\;
\frac{281}{144}\;s_1^3\;-\;\frac{29695}{864}\;s_2\;-\;
\frac{767}{72}\;s_1\;s_2\;+\;
\frac{1383}{32}\; s_3,
\label{m}
\\
\nn\\
&&
m_{4,2}=-\frac{143}{12}\;+\;\frac{5}4\; s_1,\quad
m_{4,3}=-\frac{19}{144}\;+\;\frac{1}{48} \;s_1.\nn
\ena

Then for the coefficients in (\ref{Mamunu}) we have:
\bea
&&
M^{(3,0)}=
\begin{pmatrix}
-\frac5{24}+\frac{5}{16}s'_1\\
\\
-\frac5{24}+\frac{5}{32}{s'_1}^2\\
\\
\frac{65}{48}+\frac{625}{96}s'_1-\frac{95}{24}{s'_1}^2
-\frac{5}{32}{s'_1}^3\\
\\
-\frac{1685}{144}+\frac{1735}{288}s'_1+\frac{505}{144}{s'_1}^2
+\frac{1145}{864}{s'_1}^3+\frac{55}{144}{s'_1}^4	
\end{pmatrix}\otimes
\label{M30}\\
\nn\\
&&
\begin{pmatrix}
2304-48 s'_1 - 12 {s'_1}^2-20{s'_1}^3+3 {s'_1}^4,\;\;&
24(172-64 s'_1+15 {s'_1}^2),\;\;&
6(-8+{s'_1}^2),\;\; -72
\end{pmatrix}
\nn\\
\nn\\
&&
M^{(3,1)}=\sum_{j=0}^7 {s'_1}^j M^{(3,1,j)}
\label{M31}\\
&&
 M^{(3,1,0)}=
\begin{pmatrix}
-\frac{976}3&-\frac{1193}3&\frac{41}2&\frac{51}4\\
\\
-\frac{1732}3&-\frac{1709}3&\frac{33}2&\frac{63}4\\
\\
-\frac{39496}3&-\frac{41675}6&\frac{1345}{12}&\frac{1417}8\\
\\
-\frac{137576}9&-\frac{28699}6&\frac{26227}{36}&\frac{10027}{24}
\end{pmatrix},
\quad
M^{(3,1,1)}=
\begin{pmatrix}
-\frac{1705}6&-36&-9&0\\
\\
-\frac{373}6&-\frac{327}2&\frac{33}4&\frac{63}8\\
\\
\frac{203317}{12}&16953&-436&-\frac{1599}4\\
\\
-\frac{365713}{36}&-\frac{292466}9&\frac{5849}{18}&\frac{1613}3
\end{pmatrix},
\nn\\
\nn\\
\nn\\
&&
M^{(3,1,2)}=
\begin{pmatrix}
	\frac{2059}{24}&-\frac{255}4&-\frac{17}{16}&0\\
	\\
	-\frac{5003}{24}&-\frac{163}4&-\frac{93}{16}&0\\
	\\
	\frac{13075}{16}&-\frac{115843}{24}&\frac{6527}{96}&-\frac{51}8\\
	\\
	-\frac{3085213}{144}&-\frac{399049}{72}&\frac{3437}{288}&
	\frac{2169}8
\end{pmatrix},\quad
M^{(3,1,3)}=
\begin{pmatrix}
	-\frac{83}{24}&0&0&0\\
	\\
	\frac{915}{16}&-\frac{315}8&-\frac{21}{32}&0\\
	\\
	-\frac{61081}{144}&\frac{8067}{4}&\frac{605}{16}&0\\
	\\
	\frac{739321}{144}&\frac{13460}{27}&-\frac{619}{12}&
	\frac{73}{12}
\end{pmatrix},
\nn\\
\nn\\
\nn\\
\nn\\
&&
M^{(3,1,4)}=
\begin{pmatrix}
	-\frac{17}{32}&0&0&0\\
	\\
	-\frac{63}{32}&0&0&0\\
	\\
	-\frac{15805}{192}&\frac{255}{8}&\frac{17}{32}&0\\
	\\
	-\frac{1897373}{1728}&-\frac{34807}{24}&-\frac{1139}{32}&0
\end{pmatrix},\quad
M^{(3,1,5)}=
\begin{pmatrix}
0 & 0 & 0 & 0\\
\\
-\frac{21}{64} & 0&0 &0\\
\\
\frac{1765}{96} & 0 & 0 &0\\
\\
\frac{103159}{432} & -\frac{365}{12} &-\frac{73}{144}&0
\end{pmatrix}
\nn\\
\nn\\
\nn\\
\nn\\
&&
M^{(3,1,6)}=
\begin{pmatrix}
0 & 0 & 0 & 0\\
\\
0 & 0 & 0 & 0\\
\\
\frac{17}{64} & 0 & 0& 0\\
\\
-\frac{33113}{1728} & 0 & 0& 0
\end{pmatrix},\quad\quad\quad
M^{(3,1,7)}=
\begin{pmatrix}
	0 & 0 & 0 & 0\\
\\
	0 & 0 & 0 & 0\\
\\
	0 & 0 & 0& 0\\
\\
-\frac{73}{288} & 0 & 0& 0
\end{pmatrix}
\nn\\
\nn\\
\nn\\
&&
M^{(3,2)}=\sum_{j=0}^6 {s'_1}^j M^{(3,2,j)}
\label{M32}\\
&&
M^{(3,2,0)}=
\begin{pmatrix}
\frac{383}3 & \frac{490}3 & \frac92 & 0\\
\\
97 & 256 & \frac12 & -\frac32\\
\\
-\frac{130265}{18} & -\frac{25475}{9} & \frac{2689}{12} & 140\\
\\
\frac{1032469}{54} &
\frac{748006}{27} &
-\frac{5165}{12} &
-\frac{1913}4
\end{pmatrix},\quad
M^{(3,2,1)}=
\begin{pmatrix}
-\frac{116}3 & 0 & 0 & 0\\
\\
\frac{295}2 & 65 & \frac{15}4 & 0\\
\\
-\frac{14099}9 & -\frac{8498}3 & -\frac{285}2 & 0\\
\\
\frac{196009}{18} & \frac{32920}{27} & \frac{3347}{18}
& -\frac{443}6
\end{pmatrix}
\nn\\
\nn\\
\nn\\
&&
M^{(3,2,2)}=
\begin{pmatrix}
\frac{11}3 & 0 & 0 & 0\\
\\
-\frac{179}4 & \frac{15}2 & \frac18 & 0\\
\\
-\frac{6509}9 & -\frac{2345}3 & -\frac{167}{12} & 0\\
\\
-\frac{153319}{72} & \frac{154157}{36} & \frac{1863}{16} &
\frac{11}4	
\end{pmatrix},
\quad
M^{(3,2,3)}=
\begin{pmatrix}
0 & 0 & 0 & 0\\
\\
\frac{7}3 & 0 & 0 & 0\\
\\
-\frac{517}9 & 0 & 0 & 0\\
\\
\frac{27587}{72} & \frac{23239}{54} & \frac{947}{72} & 0	
\end{pmatrix},
\nn\\
\nn\\
\nn\\
&&
M^{(3,2,4)}=
\begin{pmatrix}
0 & 0 & 0 & 0\\
\\
\frac1{16} & 0 & 0 & 0\\
\\
-\frac{23}{3} & 0 & 0 & 0\\
\\
-\frac{96763}{864} & -\frac{55}4 & -\frac{11}{48} & 0
\end{pmatrix},\quad
M^{(3,2,5)}=
\begin{pmatrix}
	0 & 0 & 0 & 0\\
	\\
	0 & 0 & 0 & 0\\
\\
	0 & 0 & 0 & 0\\
\\
	\frac{3463}{432} & 0 & 0 & 0
\end{pmatrix},\quad
M^{(3,2,6)}=
\begin{pmatrix}
	0 & 0 & 0 & 0\\
	\\
	0 & 0 & 0 & 0\\
	\\
	0 & 0 & 0 & 0\\
	\\
	-\frac{1}{96} & 0 & 0 & 0
\end{pmatrix}
\nn\\
\nn\\
\nn\\
&&
M^{(3,3)}=
\begin{pmatrix}
-1 & 0 & 0 & 0\\
\\
-\frac{128}3+\frac{41}3 s'_1-\frac{13}{24} {s'_1}^2 & -\frac{55}3 & -\frac34 & 0	\\
\\
\frac{13237}{18}-\frac{1667}9 s'_1+\frac{721}{18} {s'_1}^2 & \frac{17056}9 & 48 & 0	\\
\\
-\frac{79651}{18}+\frac{481}{54} s'_1+\frac{13067}{216} {s'_1}^2+\frac{1169}{36} {s'_1}^3 & -\frac{54623}9-\frac{22418}{27}s'_1+\frac{238}9 {s'_1}^2 &
-\frac{4525}{36}-\frac{215}6 s'_1-\frac{11}{18} {s'_1}^2 & \frac{13}3
\end{pmatrix}\nn\\
\label{M33}\\
\nn\\
&&
M^{(3,4)}=
\begin{pmatrix}
0 & 0 & 0 & 0\\
\\
-1 & 0 & 0 & 0\\
\\
-\frac{112}3 & 0 & 0 & 0\\
\\
\frac{11068}{27}-\frac{3256}{27}s'_1
-\frac{140}{27}{s'_1}^2 & \frac{1160}{27} & \frac{14}3& 0
\end{pmatrix},\quad
M^{(3,5)}=
\begin{pmatrix}
0 & 0 & 0 & 0\\
\\
0 & 0 & 0 & 0\\
\\
0 & 0 & 0 & 0\\
\\
\frac{140}{9} & 0 & 0 & 0
\end{pmatrix}
\label{M34M35}
\ena

Further
\bea
&&
M^{(1,0)}=(s'_1-2)
\begin{pmatrix}
	\frac{5}{48}\\
	\\
	-\frac1{96}(s'_1+2)\\
	\\
	-\frac{1}{288}(-933-94 s'_1+15 {s'_1}^2)\\
	\\
	-\frac{1}{864}(757+304 s'_1- 243 {s'_1}^2+34 {s'_1}^3)
\end{pmatrix}\otimes
\label{M10}\\
\nn\\
&&
\begin{pmatrix}
	1792-272 s'_1 + 196 {s'_1}^2-28{s'_1}^3+3 {s'_1}^4,\;\;\;\;&
	8(312-116 s'_1+45 {s'_1}^2),\;\;\;\;&
	6(-4-4s'_1+{s'_1}^2),\;\;\;\;& -72
\end{pmatrix}
\nn\\
\nn\\
\nn\\
&&
M^{(1,1)}=\sum_{j=0}^7 {s'_1}^j M^{(1,1,j)}
\label{M11}\\
&&
M^{(1,1,0)}=
\begin{pmatrix}
	-\frac{286}3&\frac{1030}3&\frac{39}4&-\frac{3}4\\
	\\
	180&\frac{784}3&-3&-9\\
	\\
	-\frac{41542}9&\frac{9625}3&\frac{6511}{24}&\frac{1159}8\\
	\\
	\frac{144454}{27}&\frac{165101}{27}&-\frac{2749}{24}&-\frac{5647}{24}
\end{pmatrix},
\quad
M^{(1,1,1)}=
\begin{pmatrix}
	\frac{33}2&-\frac{409}3&-\frac{13}4&0\\
	\\
	-\frac{422}3&-175&-\frac{1}8&\frac{21}8\\
	\\
	\frac{9347}{4}&-\frac{7517}{18}&-\frac{1411}{24}&-\frac{119}4\\
	\\
	-\frac{471955}{108}&-\frac{122885}{18}&-\frac{991}{72}&\frac{745}6
\end{pmatrix},
\nn\\
\nn\\
\nn\\
&&
M^{(1,1,2)}=
\begin{pmatrix}
	\frac{59}{8}&\frac{15}4&\frac{1}{16}&0\\
	\\
	\frac{821}{12}&\frac{167}2&\frac{9}{8}&0\\
	\\
	-\frac{138575}{144}&-\frac{11875}{8}&-\frac{2099}{96}&\frac{3}8\\
	\\
	\frac{364933}{144}&\frac{138691}{24}&\frac{19939}{288}&
	-\frac{1303}{24}
\end{pmatrix},\quad
M^{(1,1,3)}=
\begin{pmatrix}
	-\frac{21}{8}&0&0&0\\
	\\
	-\frac{751}{48}&-\frac{105}8&-\frac{7}{32}&0\\
	\\
	\frac{6817}{48}&\frac{2603}{12}&\frac{197}{48}&0\\
	\\
	-\frac{167467}{144}&-\frac{55436}{27}&-\frac{317}{12}&
	\frac{139}{12}
\end{pmatrix},
\nn\\
\nn\\
\nn\\
\nn\\
&&
M^{(1,1,4)}=
\begin{pmatrix}
	\frac{1}{32}&0&0&0\\
	\\
	\frac{59}{48}&0&0&0\\
	\\
	-\frac{12329}{576}&-\frac{15}{8}&-\frac{1}{32}&0\\
	\\
	\frac{708185}{1728}&\frac{96035}{216}&\frac{215}{32}&0
\end{pmatrix},\quad
M^{(1,1,5)}=
\begin{pmatrix}
	0 & 0 & 0 & 0\\
	\\
	-\frac{7}{64} & 0&0 &0\\
	\\
	\frac{245}{96} & 0 & 0 &0\\
	\\
	-\frac{18269}{216} & -\frac{695}{12} &-\frac{139}{144}&0
\end{pmatrix}
\nn\\
\nn\\
\nn\\
\nn\\
&&
M^{(1,1,6)}=
\begin{pmatrix}
	0 & 0 & 0 & 0\\
	\\
	0 & 0 & 0 & 0\\
	\\
	-\frac{1}{64} & 0 & 0& 0\\
	\\
	\frac{3599}{576} & 0 & 0& 0
\end{pmatrix},\quad\quad\quad
M^{(1,1,7)}=
\begin{pmatrix}
	0 & 0 & 0 & 0\\
	\\
	0 & 0 & 0 & 0\\
	\\
	0 & 0 & 0& 0\\
	\\
	-\frac{139}{288} & 0 & 0& 0
\end{pmatrix}
\nn\\
\nn\\
&&
M^{(1,2)}=\sum_{j=0}^6 {s'_1}^j M^{(1,2,j)}
\label{M12}\\
&&
M^{(1,2,0)}=
\begin{pmatrix}
	-\frac{187}3 & -\frac{14}3 & \frac12 & 0\\
	\\
	-\frac{62}2 & -84 & -\frac52 & -\frac32\\
	\\
	-\frac{17987}{18} & -\frac{17809}{9} & \frac{235}{4} & 14\\
	\\
	-\frac{67313}{54} &
	-\frac{201259}{27} &
	-\frac{875}{9} &
	\frac{1009}{12}
\end{pmatrix},\quad
M^{(1,2,1)}=
\begin{pmatrix}
	16 & 0 & 0 & 0\\
	\\
	\frac{29}6 & 19 & \frac{1}4 & 0\\
	\\
	\frac{3047}9 & -\frac{1150}3 & -\frac{37}6 & 0\\
	\\
	\frac{26317}{18} & 3873 & \frac{185}{4}
	& -\frac{161}6
\end{pmatrix}
\nn\\
\nn\\
\nn\\
&&
M^{(1,2,2)}=
\begin{pmatrix}
	\frac{1}6 & 0 & 0 & 0\\
	\\
	\frac{65}{12}& \frac{15}2 & \frac18 & 0\\
	\\
	\frac{413}{12} & -\frac{203}3 & -\frac{17}{12} & 0\\
	\\
	-\frac{8169}{8} & -\frac{46219}{36} & -\frac{3305}{144} &
	-\frac{25}4	
\end{pmatrix},
\quad
M^{(1,2,3)}=
\begin{pmatrix}
	0 & 0 & 0 & 0\\
	\\
	\frac{1}{12} & 0 & 0 & 0\\
	\\
	-\frac{203}{18} & 0 & 0 & 0\\
	\\
	\frac{54263}{216} & \frac{11503}{54} & \frac{235}{72} & 0	
\end{pmatrix},
\nn\\
\nn\\
\nn\\
&&
M^{(1,2,4)}=
\begin{pmatrix}
	0 & 0 & 0 & 0\\
	\\
	\frac1{16} & 0 & 0 & 0\\
	\\
	-\frac{2}{3} & 0 & 0 & 0\\
	\\
	\frac{7801}{864} & \frac{125}4 & \frac{25}{48} & 0
\end{pmatrix},\quad
M^{(1,2,5)}=
\begin{pmatrix}
	0 & 0 & 0 & 0\\
	\\
	0 & 0 & 0 & 0\\
	\\
	0 & 0 & 0 & 0\\
	\\
	\frac{661}{432} & 0 & 0 & 0
\end{pmatrix},\quad
M^{(1,2,6)}=
\begin{pmatrix}
	0 & 0 & 0 & 0\\
	\\
	0 & 0 & 0 & 0\\
	\\
	0 & 0 & 0 & 0\\
	\\
	\frac{25}{96} & 0 & 0 & 0
\end{pmatrix}
\nn\\
\nn\\
\nn\\
&&
M^{(1,3)}=
\begin{pmatrix}
	-\frac{11}3 & 0 & 0 & 0\\
	\\
	-\frac{47}6-\frac{5}6 s'_1-\frac{7}{24} {s'_1}^2 & -\frac{61}3 & -\frac14 & 0	\\
	\\
	-\frac{899}{18}-\frac{149}9 s'_1+\frac{85}{18} {s'_1}^2 & \frac{2080}9 & \frac83 & 0	\\
	\\
	\frac{17023}{27}-\frac{1376}{9} s'_1-\frac{2435}{24} {s'_1}^2-\frac{257}{36} {s'_1}^3-\frac{23}{12}{s'_1}^4 & \frac{15329}{27}-\frac{1322}{9}s'_1-\frac{1472}9 {s'_1}^2 &
	\frac{227}{12}-\frac{35}{18} s'_1-\frac{22}{9} {s'_1}^2 & \frac{49}3
\end{pmatrix}\nn\\
\label{M13}\\
\nn\\
&&
M^{(1,4)}=
\begin{pmatrix}
	0 & 0 & 0 & 0\\
	\\
	\frac13 & 0 & 0 & 0\\
	\\
	\frac{40}9 & 0 & 0 & 0\\
	\\
	\frac{2350}{27}+\frac{580}{27}s'_1
	+\frac{154}{27}{s'_1}^2 & \frac{5672}{27} & \frac{26}9& 0
\end{pmatrix},\quad
M^{(1,5)}=
\begin{pmatrix}
	0 & 0 & 0 & 0\\
	\\
	0 & 0 & 0 & 0\\
	\\
	0 & 0 & 0 & 0\\
	\\
	-\frac{212}{27} & 0 & 0 & 0
\end{pmatrix}
\label{M14M15}
\ena

And finally we have:
\bea
&&
M^{(-1,0)}=(s'_1+2)
\begin{pmatrix}
	\frac{-5}{48}\\
	\\
	-\frac1{96}(s'_1-2)\\
	\\
	-\frac{1}{288}(-933+94 s'_1+15 {s'_1}^2)\\
	\\
	-\frac{1}{864}(737+324 s'_1+ 223 {s'_1}^2+14 {s'_1}^3)
\end{pmatrix}\otimes
\label{Mm10}\\
\nn\\
&&
\begin{pmatrix}
	896+1776 s'_1 - 348 {s'_1}^2-36{s'_1}^3+3 {s'_1}^4,\;\;\;\;&
	8(468-40 s'_1+45 {s'_1}^2),\;\;\;\;&
	6(8-8s'_1+{s'_1}^2),\;\;\;\;& -72
\end{pmatrix}
\nn\\
\nn\\
\nn\\
&&
M^{(-1,1)}=\sum_{j=0}^7 {s'_1}^j M^{(-1,1,j)}
\label{Mm11}\\
&&
M^{(-1,1,0)}=
\begin{pmatrix}
	1372&\frac{2359}3&-\frac{13}2&-\frac{57}4\\
	\\
	-\frac{1396}3&-\frac{1667}3&\frac12&-\frac{45}4\\
	\\
	\frac{346954}9&\frac{120157}6&-\frac{1453}{12}&-\frac{2675}8\\
	\\
	-\frac{276418}{27}&-\frac{216115}{54}&\frac{2369}{36}&\frac{533}{8}
\end{pmatrix},
\quad
M^{(-1,1,1)}=
\begin{pmatrix}
	-\frac{373}2&\frac{262}3&-\frac{13}2&0\\
	\\
	\frac{767}6&\frac{341}2&\frac{9}4&-\frac{21}8\\
	\\
	-\frac{239407}{36}&\frac{64798}{9}&-\frac{2207}{12}&-\frac{295}4\\
	\\
	\frac{707131}{108}&\frac{22145}{27}&\frac{593}{36}&-\frac{88}3
\end{pmatrix},
\nn\\
\nn\\
\nn\\
&&
M^{(-1,1,2)}=
\begin{pmatrix}
	-\frac{1517}{24}&\frac{285}4&\frac{19}{16}&0\\
	\\
	-\frac{1205}{24}&-\frac{269}4&-\frac{19}{16}&0\\
	\\
	-\frac{36509}{16}&\frac{433}{24}&\frac{1051}{96}&\frac{57}8\\
	\\
	\frac{928097}{432}&\frac{2923}{72}&-\frac{12407}{288}&
	\frac{247}{24}
\end{pmatrix},\quad
M^{(-1,1,3)}=
\begin{pmatrix}
	-\frac{115}{24}&0&0&0\\
	\\
	\frac{375}{16}&\frac{105}8&\frac{7}{32}&0\\
	\\
	\frac{85999}{144}&\frac{3901}{12}&\frac{451}{48}&0\\
	\\
	-\frac{267317}{432}&\frac{14296}{9}&-\frac{233}{36}&
	-\frac{227}{12}
\end{pmatrix},
\nn\\
\nn\\
\nn\\
\nn\\
&&
M^{(-1,1,4)}=
\begin{pmatrix}
	\frac{19}{32}&0&0&0\\
	\\
	-\frac{91}{96}&0&0&0\\
	\\
	\frac{11837}{576}&-\frac{285}{8}&-\frac{19}{32}&0\\
	\\
	-\frac{612025}{1728}&-\frac{48059}{216}&-\frac{365}{96}&0
\end{pmatrix},\quad
M^{(-1,1,5)}=
\begin{pmatrix}
	0 & 0 & 0 & 0\\
	\\
	\frac{7}{64} & 0&0 &0\\
	\\
	\frac{175}{32} & 0 & 0 &0\\
	\\
	\frac{67565}{432} & \frac{1135}{12} &\frac{227}{144}&0
\end{pmatrix}
\nn\\
\nn\\
\nn\\
\nn\\
&&
M^{(-1,1,6)}=
\begin{pmatrix}
	0 & 0 & 0 & 0\\
	\\
	0 & 0 & 0 & 0\\
	\\
	-\frac{19}{64} & 0 & 0& 0\\
	\\
	-\frac{8509}{1728} & 0 & 0& 0
\end{pmatrix},\quad\quad\quad
M^{(-1,1,7)}=
\begin{pmatrix}
	0 & 0 & 0 & 0\\
	\\
	0 & 0 & 0 & 0\\
	\\
	0 & 0 & 0& 0\\
	\\
	\frac{227}{288} & 0 & 0& 0
\end{pmatrix}
\nn\\
\nn\\
&&
M^{(-1,2)}=\sum_{j=0}^6 {s'_1}^j M^{(-1,2,j)}
\label{Mm12}\\
&&
M^{(-1,2,0)}=
\begin{pmatrix}
	-51& -\frac{662}3 & \frac12 & 0\\
	\\
	403 & 196 & -\frac12 & -\frac32\\
	\\
	-\frac{21907}{6} & -\frac{93515}{9} & \frac{865}{12} & 104\\
	\\
	\frac{79429}{54} &
	\frac{65054}{9} &
	-\frac{595}{36} &
	-\frac{1555}{12}
\end{pmatrix},\quad
M^{(-1,2,1)}=
\begin{pmatrix}
	\frac{296}3 & 0 & 0 & 0\\
	\\
	-\frac{887}{6} & -19 & -\frac{5}4 & 0\\
	\\
	\frac{3845}3 & -142 & \frac{157}6 & 0\\
	\\
	\frac{1679}{18} & -4498 & -\frac{22}{3}
	& \frac{337}6
\end{pmatrix}
\nn\\
\nn\\
\nn\\
&&
M^{(-1,2,2)}=
\begin{pmatrix}
	-\frac{4}3 & 0 & 0 & 0\\
	\\
	-\frac{133}{12}& \frac{15}2 & \frac18 & 0\\
	\\
	-\frac{1979}{6} & -\frac{1229}3 & -\frac{107}{12} & 0\\
	\\
	\frac{55963}{24} & \frac{13303}{24} & \frac{317}{144} &
	\frac{11}4	
\end{pmatrix},
\quad
M^{(-1,2,3)}=
\begin{pmatrix}
	0 & 0 & 0 & 0\\
	\\
	-\frac{7}{6} & 0 & 0 & 0\\
	\\
	-\frac{244}{9} & 0 & 0 & 0\\
	\\
	-\frac{241099}{216} & -\frac{20465}{54} & -\frac{581}{72} & 0	
\end{pmatrix},
\nn\\
\nn\\
\nn\\
&&
M^{(-1,2,4)}=
\begin{pmatrix}
	0 & 0 & 0 & 0\\
	\\
	\frac1{16} & 0 & 0 & 0\\
	\\
	-\frac{11}{3} & 0 & 0 & 0\\
	\\
	-\frac{2233}{32} & -\frac{55}4 & -\frac{1}{48} & 0
\end{pmatrix},\quad
M^{(-1,2,5)}=
\begin{pmatrix}
	0 & 0 & 0 & 0\\
	\\
	0 & 0 & 0 & 0\\
	\\
	0 & 0 & 0 & 0\\
	\\
	-\frac{2453}{432} & 0 & 0 & 0
\end{pmatrix},\quad
M^{(-1,2,6)}=
\begin{pmatrix}
	0 & 0 & 0 & 0\\
	\\
	0 & 0 & 0 & 0\\
	\\
	0 & 0 & 0 & 0\\
	\\
	-\frac{11}{96} & 0 & 0 & 0
\end{pmatrix}
\nn\\
\nn\\
\nn\\
&&
M^{(-1,3)}=\label{Mm13}\\
&&
\begin{pmatrix}
	-\frac{83}3 & 0 & 0 & 0\\
	\\
	\frac{94}3+\frac{56}3 s'_1-\frac{1}{24} {s'_1}^2 & -\frac{91}3 & -\frac14 & 0	\\
	\\
	\frac{3055}{6}-\frac{2519}9 s'_1+\frac{481}{18} {s'_1}^2 & \frac{12448}9 & \frac83 & 0	\\
	\\
	-\frac{222865}{54}+\frac{113927}{54} s'_1+
	\frac{37037}{72} {s'_1}^2+\frac{7327}{108} {s'_1}^3+\frac{11}{6}{s'_1}^4 & -\frac{8153}{27}+\frac{5270}{27}s'_1+\frac{778}9 {s'_1}^2 &
	-\frac{211}{12}+\frac{127}{18} s'_1+\frac{55}{18} {s'_1}^2 & -\frac{59}3
\end{pmatrix}
\nn\\
\nn\\
\nn\\
&&
M^{(-1,4)}=
\begin{pmatrix}
	0 & 0 & 0 & 0\\
	\\
	-\frac{19}3 & 0 & 0 & 0\\
	\\
	\frac{832}9 & 0 & 0 & 0\\
	\\
	-\frac{21424}{27}-\frac{4208}{27}s'_1
	-\frac{680}{27}{s'_1}^2 & -\frac{3448}{27} & -\frac{38}9& 0
\end{pmatrix},\quad
M^{(-1,5)}=
\begin{pmatrix}
	0 & 0 & 0 & 0\\
	\\
	0 & 0 & 0 & 0\\
	\\
	0 & 0 & 0 & 0\\
	\\
	\frac{1268}{27} & 0 & 0 & 0
\end{pmatrix}
\label{Mm14M15}
\ena

\end{document}